\documentclass[aps,prd,superscriptaddress,twocolumn,nofootinbib,amsmath,amssymb,nolongbibliography]{revtex4-2}
\usepackage{graphicx}
\usepackage[colorlinks=true]{hyperref}
\usepackage{url}

\usepackage{color}

\begin{document}

\title{Effects of mirror birefringence and its fluctuations to \\laser interferometric gravitational wave detectors}

\author{Yuta~Michimura}
\email{yuta@caltech.edu}
\affiliation{LIGO Laboratory, California Institute of Technology, Pasadena, California 91125, USA}
\affiliation{Research Center for the Early Universe (RESCEU), Graduate School of Science, University of Tokyo, Tokyo 113-0033, Japan}
\affiliation{PRESTO, Japan Science and Technology Agency (JST), Kawaguchi, Saitama 332-0012, Japan}
\author{Haoyu~Wang}
\affiliation{Research Center for the Early Universe (RESCEU), Graduate School of Science, University of Tokyo, Tokyo 113-0033, Japan}
\author{Francisco Salces--Carcoba}
\affiliation{LIGO Laboratory, California Institute of Technology, Pasadena, California 91125, USA}
\author{Christopher~Wipf}
\affiliation{LIGO Laboratory, California Institute of Technology, Pasadena, California 91125, USA}
\author{Aidan~Brooks}
\affiliation{LIGO Laboratory, California Institute of Technology, Pasadena, California 91125, USA}
\author{Koji~Arai}
\affiliation{LIGO Laboratory, California Institute of Technology, Pasadena, California 91125, USA}
\author{Rana~X~Adhikari}
\affiliation{LIGO Laboratory, California Institute of Technology, Pasadena, California 91125, USA}

\date{\today}

\begin{abstract}
Crystalline materials are promising candidates as substrates or high-reflective coatings of mirrors to reduce thermal noises in future laser interferometric gravitational wave detectors. However, birefringence of such materials could degrade the sensitivity of gravitational wave detectors, not only because it can introduce optical losses, but also because its fluctuations create extra phase noise in the arm cavity reflected beam. In this paper, we analytically estimate the effects of birefringence and its fluctuations in the mirror substrate and coating for gravitational wave detectors. Our calculations show that the requirements for the birefringence fluctuations in silicon substrate and AlGaAs coating will be on the order of $10^{-8}$ and  $10^{-10}$ rad/$\sqrt{\rm Hz}$ at 100~Hz, respectively, for future gravitational wave detectors. We also point out that optical cavity response needs to be carefully taken into account to estimate optical losses from depolarization.
\end{abstract}

\maketitle

\section{Introduction}
The first detections of gravitational waves from binary black holes~\cite{GW150914} and binary neutron stars~\cite{GW170817,Multimessenger} by Advanced LIGO~\cite{aLIGO} and Advanced Virgo~\cite{AdV} inaugurated gravitational wave physics and astronomy. Improvements in the sensitivity of these laser interferometric detectors in recent years enabled routine detections and more precise binary parameter estimation~\cite{ObservingScenarioPaper}. Further improvements in the astrophysical reach of these detectors will allow us to study the origin of massive black holes, the neutron star equation of state, alternative gravity theories, and cosmology.

The fundamental limitation to the sensitivity of these detectors at the most sensitive frequency band is set by thermal vibrations of mirror surface~\cite{RanaRMP}. KAGRA~\cite{AsoKAGRA,PTEP01KAGRA} and other concepts of future gravitational wave detectors plan to utilize cryogenic crystalline test mass mirrors for thermal noise reduction, instead of fused silica mirrors at room temperature. KAGRA uses sapphire test masses, and plan to cool them down to 22~K~\cite{PSOKAGRA}. Voyager is an upgrade plan of LIGO to use 123~K silicon to increase the astrophysical reach by a factor of 4--5 over Advanced LIGO design~\cite{Voyager}. The next generation detectors such as Einstein Telescope~\cite{ET,ET-0007B-20} also plan to use silicon test masses at cryogenic temperatures for the low frequency detectors, and Cosmic Explorer~\cite{CE,CEHorizon} considers using them for an upgrade. In addition, crystalline coatings such as AlGaAs coating~\cite{Cole2013} and AlGaP coating~\cite{AlGaP} are considered as promising candidates to reduce coating Brownian noise, instead of amorphous silica and tantala coating.

Although crystalline materials are promising to reduce thermal noise, it has been pointed out that slight birefringence of mirror substrates and coatings could cause optical losses due to depolarization of the light, and cause degradation of interferometric contrast~\cite{Winkler1994}. The birefringence and its inhomogeneity of sapphire input test masses of KAGRA were found to be higher than expected~\cite{SapphireInhomogeneity,HiroseSapphire}, and around 10\% of power was lost on reflection due to depolarization, when arm cavities are not on resonance~\cite{PTEP01KAGRA}. Ideally, crystalline silicon is a cubic crystal and optically isotropic, but could have strain-induced birefringence from crystal dislocations and due to support in the mirror suspension system. Birefringence measurements in silicon mirrors have revealed that the amount of the static birefringence is $\Delta n \sim 10^{-7}$ or less at laser wavelengths of 1.55~\cite{AEISiliconBirefringence} and 2~$\mu$m~\cite{UWASiliconBirefringence} at room temperature, which satisfies the optical loss requirements for future detectors. Also, previous cavity experiments using AlGaAs coatings reported birefringence at 1~mrad level~\cite{Cole2013,Winkler2021,Tanioka2023}.

These past studies have focused on the static birefringence and optical losses from the depolarization. However, recent measurement of thermal noises in crystalline mirror coatings at cryogenic temperatures reported excess birefringent noise, which could limit the sensitivity of future gravitational wave detectors~\cite{JILA2022}. Theoretical calculations on thermal fluctuations of birefringence in crystalline mirror coatings have also revealed that the noise from these fluctuations could be similar to Brownian noise~\cite{ThermorefringentNoise}. It is also worth noting that experiments to search for vacuum magnetic birefringence, such as PVLAS (Polarizzazione del Vuoto con LASer) and OVAL (Observing VAcuum with Laser), have been suspected to be limited by thermal birefringence noise of mirrors~\cite{Bielsa2009,Hollis2019,Zavattini2018,Ejlli2020,KamiokaDron}. These temporal birefringence fluctuations could also limit optical cavity based axion dark matter searches using the birefringence effect from axion-photon coupling~\cite{DANCE,ADBC,LIDA,ADAM-GD,ADAM-GDTR}.

In this paper, we study the effects of birefringence and its fluctuations to gravitational wave detectors based on the Fabry-P{\'e}rot-Michelson interferometer. We show that the polarization axis and the crystal axes of arm cavity mirrors need to be aligned to avoid optical losses and to reduce noises from birefringence fluctuations. We also show that the cavity response to birefringence needs to be correctly taken into account for estimating the noises and the optical losses of arm cavities. We start by analytically describing the cavity response to birefringence in Sec.~\ref{CavityResponse}. In Sec.~\ref{Noises}, we focus on noises from substrate birefringence and coating birefringence, and derive requirements for their fluctuations for future gravitational wave detectors. In Sec.~\ref{PowerLosses}, we expand our formulation to include spatial higher order modes, and discuss power losses from inhomogeneous birefringence of the substrate and the coating. Our conclusions and outlook are summarized in Sec.~\ref{Conclusion}. 

Throughout the paper, we use 0.1\% as a requirement threshold for the optical losses from polarization. In this way the optical losses from polarization will be small enough, as future gravitational wave detector designs require total optical loss to be less than 10\%~\cite{Oelker2016}.

\section{Cavity response to birefringence} \label{CavityResponse}
\begin{figure}
\begin{center}
\includegraphics[width=\hsize]{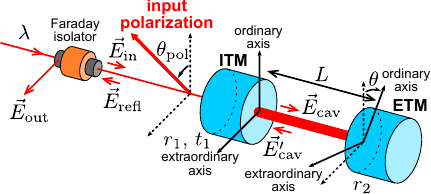}
\end{center}
\caption{\label{FPcavity} The schematic of a Fabry-P{\'e}rot cavity with mirror crystal axes and input beam polarization axis illustrated. With respect to the ITM ordinary axis, the input polarization is rotated by $\theta_{\rm pol}$ and the ETM ordinary axis is rotated by $\theta$.}
\end{figure}

Let us consider a Fabry-P{\'e}rot cavity formed by an input test mass (ITM) and an end test mass (ETM) mirrors as shown in Fig.~\ref{FPcavity}. We consider birefringence of ITM substrate, ITM high-reflective coating, and ETM high-reflective coating. The ordinary axis of the ETM coating is rotated by $\theta$ with respect to that of ITM. The input beam is linearly polarized and its polarization is rotated by $\theta_{\rm pol}$ with respect to the ordinary axis of ITM. We assume that the crystal axes of ITM substrate are aligned with those of its coating. This will not affect the results of this paper, as we will treat the substrate birefringence and the coating birefringence independently in the following sections.

For calculating the cavity response to birefringence, we can use the Jones matrix formalism~\cite{JonesCalculus}. In the basis of ITM crystal axes, the electric field of the input beam can be written as
\begin{equation}
 \vec{E}_{\rm in}= (v_1 \vec{e}_{\rm o} + 
 v_2 \vec{e}_{\rm e}) E_{\rm in} =
 \begin{pmatrix}
\vec{e}_{\rm o} & \vec{e}_{\rm e} \\
\end{pmatrix}
 \vec{v}_{\rm in} E_{\rm in} ,
\end{equation}
where $\vec{e}_{\rm o}$ and $\vec{e}_{\rm e}$ are the unit vectors along with the ITM ordinary and extraordinary axes, respectively, and $\vec{v}_{\rm in} \equiv (v_1 \ v_2)^{T}$ is the unit vector representing the input polarization.

We suppose the ITM substrate is lossless, and the amplitude reflectivity and the amplitude transmission of the whole ITM is determined by the high-reflective coating. Then the amplitude transmission of ITM can be written as
\begin{equation}
 T_{\rm 1}=
 \begin{pmatrix}
t_1 & 0 \\
0 & t_1 e^{-i \frac{1}{2} \Delta \phi_{\rm t_1}} \\
\end{pmatrix} ,
\end{equation}
where $\Delta \phi_{\rm t_1}/2$ is the phase difference between the ordinary and extraordinary axes in the ITM transmission from both the substrate and the coating birefringence, and $t_1$ is the amplitude transmission of ITM. Here, we assumed that the amplitude transmission is the same for both axes. Similarly, the amplitude reflectivity of ITM and ETM from the high-reflective coating side can be written as
\begin{equation}
 R_j=
 \begin{pmatrix}
r_j & 0 \\
0 & r_j e^{-i \Delta \phi_{r_j}} \\
\end{pmatrix} ,
\end{equation}
where $\Delta \phi_{{\rm r}_j}$ is the phase difference between the ordinary and extraordinary axes in ITM and ETM reflection, and $r_j$ is the amplitude reflectivity of ITM and ETM. $j=1$ is for ITM and $j=2$ is for ETM. Also, the amplitude reflectivity of ITM from the substrate side can be written as
\begin{equation}
 S_{\rm 1}=
 \begin{pmatrix}
-r_1 & 0 \\
0 & -r_1 e^{-i \Delta \phi_{\rm s_1}} \\
\end{pmatrix} ,
\end{equation}
where $\Delta \phi_{s_1}$ is the phase difference between the ordinary and extraordinary axes in the ITM reflection from the substrate side. From the energy conservation and the time reversal symmetry, $\Delta \phi_{t_1}=\Delta \phi_{r_1}+\Delta \phi_{s_1}$. Here, we use the convention that $r_j$ and $t_1$ are real, and the sign is flipped for reflection from the ITM substrate side. We keep the coordinate axis to be the same even if the propagation direction flips on mirror reflections, so that the sign for both polarizations will be the same.

For arm cavities in gravitational wave detectors, $r_1$ and $r_2$ are designed to be $r_2 \simeq 1$, and $r_1 < r_2$, such that almost all the light is reflected back. From the phase of the cavity reflected beam, cavity length changes from gravitational waves are read out. In the following subsections, we calculate the polarization eigenmodes in the cavity, and the phase of the cavity reflected beam.

\subsection{Polarization eigenmodes in the cavity}
The electric field inside the cavity that propagates from ITM to ETM can be written as
\begin{equation} \label{eq:Ecav}
 \vec{E}_{\rm cav} = \left( I - A \right)^{-1} T_1 \vec{E}_{\rm in},
\end{equation}
with $I$ being the identity matrix. Here,
\begin{equation}
 A \equiv R_1 R(-\theta) R_2 R(\theta) e^{-i \phi} ,
\end{equation}
where $\phi=4 \pi L/\lambda$ is the phase acquired in the cavity round-trip, with $L$ and $\lambda$ being the cavity length and the laser wavelength, respectively, and
\begin{equation}
 R(\theta) \equiv
 \begin{pmatrix}
\cos{\theta} & -\sin{\theta} \\
\sin{\theta} & \cos{\theta} \\
\end{pmatrix} ,
\end{equation}
with the derivation described in Appendix~\ref{CavityDerivation}. Note that $\phi$ includes phase acquired in the ITM and ETM reflection for their ordinary axes. The resonant polarization mode is the eigenvectors of
\begin{equation}
 M_{\rm cav} \equiv \left( I - A \right)^{-1} T_1 .
\end{equation}
The cavity enhancement factors for each mode will be the eigenvalues of $M_{\rm cav}$.

When $\theta=0$, the ITM axes and the ETM axes are aligned, and the eigenvectors will be
\begin{equation} \label{theta0_eigen_vectors}
 \vec{v}_a =
 \begin{pmatrix}
1 \\
0 \\
\end{pmatrix}, \qquad
 \vec{v}_b =
 \begin{pmatrix}
0 \\
1 \\
\end{pmatrix},
\end{equation}
which means that the resonant modes are linear polarizations along the ITM ordinary axis $\vec{e}_{\rm o}$ and the extraordinary axis $\vec{e}_{\rm e}$. The cavity enhancement factors will be
\begin{equation}
 w_a = \frac{t_1}{1-r_1 r_2 e^{-i \phi}}, \qquad
 w_b = \frac{t_1 e^{-i \frac{1}{2} \Delta \phi_{\rm t_1}}}{1-r_1 r_2 e^{-i (\phi+\Delta \phi_{\rm r_1} + \Delta \phi_{\rm r_2})}} .
\end{equation}
The resonant frequency difference between two eigenmodes therefore will be
\begin{equation}  \label{resonant_freq_split_max}
 \Delta \nu = \frac{\Delta \phi_{\rm r_1} + \Delta \phi_{\rm r_2}}{ 2 \pi} \nu_{\rm FSR},
\end{equation}
where $\nu_{\rm FSR}=c/(2 L)$ is the free spectral range of the cavity.

When $\theta=\pi/2$, the ITM ordinary axis and the ETM extraordinary axis are aligned, and the eigenvectors again will be the same as the ones given in Eq.~(\ref{theta0_eigen_vectors}). The cavity enhancement factors will be
\begin{equation}
 w_a = \frac{t_1}{1-r_1 r_2 e^{-i (\phi+\Delta \phi_{\rm r_2})}}, \quad
 w_b = \frac{t_1 e^{-i \frac{1}{2} \Delta \phi_{\rm t_1}}}{1-r_1 r_2 e^{-i (\phi+\Delta \phi_{\rm r_1})}}.
\end{equation}
The resonant frequency difference between two eigenmodes therefore will be
\begin{equation} \label{resonant_freq_split_min}
 \Delta \nu = \frac{\Delta \phi_{\rm r_1} - \Delta \phi_{\rm r_2}}{ 2 \pi} \nu_{\rm FSR}.
\end{equation}

Since we defined the ITM and ETM axes such that $\Delta \phi_{{\rm r}_i}$ have the same sign for ITM and ETM, when $\theta=0$, the phase difference between the axes are added and the resonant frequency difference is maximized. When $\theta=\pi/2$, it is minimized, as the phase difference is canceled. When $0 < \theta < \pi/2$, the resonant frequency difference will be in between the maximum and the minimum.

When the resonant frequency difference is smaller than the cavity linewidth, i.e., $\Delta \phi_{{\rm r}_i} \ll 2 \pi/\mathcal{F}$, and when the effect from the ITM substrate birefringence is small, i.e., $\Delta \phi_{\rm t_1} \ll \Delta \phi_{\rm r_1} \mathcal{F}/\pi$, the resonant frequency difference can be calculated with
\begin{equation}
 \Delta \nu \simeq \frac{2 \pi (\arg{w_a} - \arg{w_b} )}{\mathcal{F}} \frac{\nu_{\rm FSR}}{ 2 \pi} ,
\end{equation}
at $\phi=0$, where
\begin{equation}
 \mathcal{F} = \frac{\pi \sqrt{r_1 r_2}}{1-r_1 r_2}
\end{equation}
is the finesse of the cavity. This can be further approximated as~\cite{Brandi1997}
\begin{equation} \label{resonant_freq_diff_theory}
 \Delta \nu \simeq \frac{\delta_{\rm EQ}}{2 \pi} \nu_{\rm FSR} ,
\end{equation}
where
\begin{equation}\label{eq:deltaEQ}
 \delta_{\rm EQ} \equiv \sqrt{(\Delta \phi_{\rm r_1}-\Delta \phi_{\rm r_2})^2+4\Delta \phi_{\rm r_1} \Delta \phi_{\rm r_2} \cos^2{\theta}},
\end{equation}
when $\delta_{\rm EQ} \ll 1$, with the derivation described in Appendix~\ref{EqPhaseDerivation}. Also, the cavity eigenmodes are linear polarizations approximated as
\begin{equation} \label{resonant_mode_theory}
 \vec{v}_a =
 \begin{pmatrix}
\cos{\theta_{\rm EQ}} \\
\sin{\theta_{\rm EQ}} \\
\end{pmatrix}, \quad
 \vec{v}_b =
 \begin{pmatrix}
-\sin{\theta_{\rm EQ}} \\
\cos{\theta_{\rm EQ}} \\
\end{pmatrix},
\end{equation}
where the polarization angle is defined by
\begin{equation} \label{thetaEQ}
 \cos{2 \theta_{\rm EQ}} = \frac{\displaystyle{\frac{\Delta \phi^{\prime}_{r_1}}{ \Delta \phi_{\rm r_2}}} + \cos{2\theta}}{\sqrt{\left( \displaystyle{\frac{\Delta \phi^{\prime}_{r_1}}{\Delta \phi_{\rm r_2}}} - 1 \right)^2+4\displaystyle{\frac{\Delta \phi^{\prime}_{r_1}}{ \Delta \phi_{\rm r_2}}} \cos^2{\theta}}} ,
\end{equation}
with
\begin{equation}
 \Delta \phi^{\prime}_{r_1} \equiv \Delta \phi_{\rm r_1} + \frac{\pi}{\mathcal{F}}\Delta \phi_{\rm t_1}  .
\end{equation}
When $\Delta \phi^{\prime}_{r_1} \gg \Delta \phi_{r_2}$, $\theta_{\rm EQ}$ is equal to zero, when $\Delta \phi^{\prime}_{r_1} = \Delta \phi_{r_2}$, $\theta_{\rm EQ}$ is equal to $\theta/2$, and when $\Delta \phi^{\prime}_{r_1} \ll \Delta \phi_{r_2}$, $\theta_{\rm EQ}$ is equal to $\theta$. Note that the polarization state resonating inside the cavity are elliptic polarizations given by $R_1 T_1 \vec{v}_{a,b} / (r_1 t_1)$, and are different from linear polarizations given by Eq.~(\ref{resonant_mode_theory}).

The mismatch between the cavity polarization mode and the input beam polarization can be calculated with
\begin{equation} \label{optical_loss}
 \Lambda^2 = 1- \left| \vec{v}_a \cdot \vec{v}_{\rm in} \right|^2 .
\end{equation}
When the input beam is linearly polarized with the polarization angle of $\theta_{\rm pol}$ such that
\begin{equation} \label{input_pol}
 \vec{v}_{\rm in} = R(\theta_{\rm pol})
 \begin{pmatrix}
1 \\
0 \\
\end{pmatrix} =
 \begin{pmatrix}
\cos{\theta_{\rm pol}} \\
\sin{\theta_{\rm pol}} \\
\end{pmatrix},
\end{equation} 
Eq.~(\ref{optical_loss}) reduces to
\begin{equation} \label{optical_loss_reduced}
 \Lambda^2 = \sin^2{(\theta_{\rm EQ}-\theta_{\rm pol})} .
\end{equation}
The mismatch will be less than than 0.1\% when $|\theta_{\rm EQ}-\theta_{\rm pol}|$ is smaller than $1.8^{\circ}$. For gravitational wave detectors, this is required for both arm cavities. This means that the axes of two arm cavities need to be aligned to the same degree. Note that mismatch do not directly mean that there is a same amount of power loss. The actual power loss also depend on the amount of birefringence, as we will discuss in Sec.~\ref{PowerLosses}.

\begin{figure}[t]
\begin{center}
\includegraphics[width=\hsize]{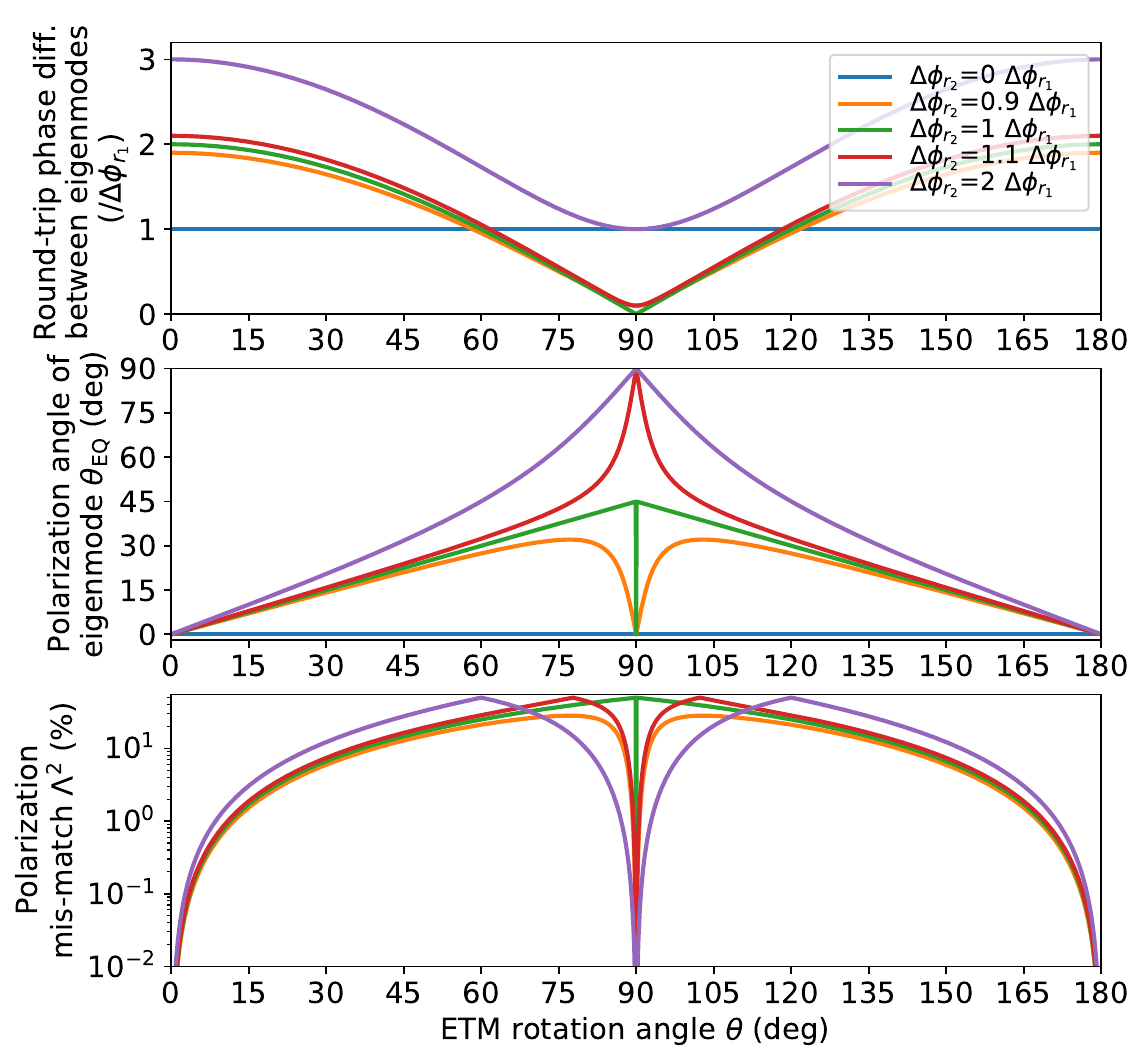}
\end{center}
\caption{\label{CavityEigenmode} The polarization eigenmodes of a Fabry-P{\'e}rot cavity as a function of ETM rotation angle $\theta$. The top panel shows the round-trip phase difference between the eigenmodes in the unit of $\Delta \phi_{\rm r_1}$, i.e., $2 \pi \Delta \nu/(\nu_{\rm FSR} \Delta \phi_{\rm r_1})$, which is proportional to the resonant frequency difference. The middle panel shows the polarization angle of the eigenmodes $\theta_{\rm EQ}$ calculated using Eq.~(\ref{thetaEQ}). The bottom panel shows the mismatch of the input beam polarization to the eigenmodes, when it is linear and aligned with ITM axes, calculated using Eq.~(\ref{optical_loss}). Different colors of the lines correspond to different $\Delta \phi_{\rm r_2}/\Delta \phi_{\rm r_1}$ ratios. Blue lines for $\Delta \phi_{\rm r_2}=0$ case in the bottom two plots are zero.}
\end{figure}

Figure~\ref{CavityEigenmode} shows the polarization eigenmodes of the cavity as a function of ETM rotation angle $\theta$, calculated using Eqs.~(\ref{resonant_freq_diff_theory}) and (\ref{thetaEQ}). As we have discussed earlier, the resonant frequency difference will be the maximized at $\theta=0$, and minimized at $\theta=\pi/2$. When $\theta=\pi/2$ and $\Delta \phi_{\rm r_1}=\Delta \phi_{\rm r_2}$, the phase difference between ordinary and extraordinary axes is completely cancelled, and two modes will be degenerate. In this case, two linear polarizations and two circular polarizations will be cavity eigenmodes, since two modes have the same resonant frequency.

The bottom panel in Fig.~\ref{CavityEigenmode} shows the mismatch calculated using Eq.~(\ref{optical_loss}), assuming the input polarization is linear and aligned with either of the ITM axes. The mismatch is nulled at $\theta=0$ and $\theta=\pi/2$. To minimize the mismatch and to make the resonant frequency difference large, aligning the ETM rotation such that $\theta=0$ and aligning the input polarization to one of the ITM axes will be the optimal choice. The requirement on the alignment will be not severe, since the dependence on the ETM rotation angle goes with $\theta^2$ at $\theta=0$.

For deriving the cavity reflected beam, we need to calculate the electric field inside the cavity that propagates from ETM to ITM. This can be written as
\begin{eqnarray}
 \vec{E}^{\prime}_{\rm cav} &=& R(-\theta) R_2 R(\theta) e^{- i \phi} M_{\rm cav} \vec{E}_{\rm in} \\
 &\equiv& M^{\prime}_{\rm cav} \vec{E}_{\rm in} . \label{eq:Ecavprime}
\end{eqnarray}
The eigenvectors of $M^{\prime}_{\rm cav}$ are the same as those of $M_{\rm cav}$ within our approximations discussed above, but the cavity enhancement factors will be slightly different. When $\theta=0$, the cavity enhancement factors will be
\begin{equation}
 w^{\prime}_a = \frac{t_1 r_2 e^{-i \phi}}{1-r_1 r_2 e^{-i \phi}} , \quad
 w^{\prime}_b = \frac{t_1 r_2 e^{-i (\phi+ \frac{1}{2} \Delta \phi_{\rm t_1}+\Delta \phi_{\rm r_2}})}{1-r_1 r_2 e^{-i (\phi+\Delta \phi_{\rm r_1} + \Delta \phi_{\rm r_2})}},
\end{equation}
and when $\theta=\pi/2$, those will be
\begin{equation}
 w^{\prime}_a = \frac{t_1 r_2 e^{-i (\phi+\Delta \phi_{\rm r_2})}}{1-r_1 r_2 e^{-i (\phi+\Delta \phi_{\rm r_2})}} , \quad
 w^{\prime}_b = \frac{t_1 r_2 e^{-i (\phi+ \frac{1}{2} \Delta \phi_{\rm t_1}})}{1-r_1 r_2 e^{-i (\phi+\Delta \phi_{\rm r_1})}}.
\end{equation}
Compared with $w_a$ and $w_b$, those have extra phase $\phi$ from the cavity round trip and extra phase $\Delta \phi_{\rm r_2}$ for the corresponding axis for one additional reflection from ETM.

\subsection{Phase of cavity reflected beam}
The noises due to temporal fluctuations of birefringence will be imprinted in the phase of the cavity reflected beam. The electric field of the cavity reflection can be written as
\begin{equation} \label{eq:Erefl}
 \vec{E}_{\rm refl} = M_{\rm refl} \vec{E}_{\rm in}
\end{equation}
where
\begin{equation}
 M_{\rm refl} \equiv S_1+
 T_1 M^{\prime}_{\rm cav} .
\end{equation}
The first term corresponds to the prompt reflection from ITM, and the second term is the ITM transmitted beam from the cavity circulating beam. In general, when the input beam polarization component is
\begin{equation}
 \vec{v}_{\rm in} = a \vec{v}^{\prime}_a + b \vec{v}^{\prime}_b ,
\end{equation}
the polarization component of the reflected beam is
\begin{equation} \label{cavity_refl}
 M_{\rm refl} \vec{v}_{\rm in} = a (S_1 + w^{\prime}_a T_1) \vec{v}^{\prime}_a + b (S_1 + w^{\prime}_b T_1) \vec{v}^{\prime}_b .
\end{equation}
Since the resonant condition of each eigenmode is generally different, it is generally $|w^{\prime}_a| \ne |w^{\prime}_b|$. Therefore, the polarization component of the cavity reflected beam will be different from the input polarization.

When we use a Faraday isolator to extract the cavity reflection, we extract the polarization component which is the same as the input polarization. Therefore, the phase of the cavity reflected beam can be calculated with
\begin{equation}
 \arg{(E_{\rm out})} = \arg{(E_{\rm refl \parallel})} = \arg{(E_{\rm in} M_{\rm refl} \vec{v}_{\rm in} \cdot \vec{v}_{\rm in})} .
\end{equation}
In the case when the input beam polarization is aligned to the ITM ordinary axis, this reflected phase is the phase of the (1,1) component of $M_{\rm refl}$, and that for the ITM extraordinary axis is the (2,2) component of $M_{\rm refl}$.

Let us first consider the effects from ITM. If we set $\Delta \phi_{\rm r_2} = 0$ and the input beam is linearly polarized with the polarization angle of $\theta_{\rm pol}$ as shown in Eq.~(\ref{input_pol}), the reflected electric field in the polarization parallel to $\vec{v}_{\rm in}$ and in the orthogonal polarization will be
\begin{eqnarray}
 \frac{E_{\rm refl \parallel}}{E_{\rm in}} &=& M_{\rm refl} \vec{v}_{\rm in} \cdot \vec{v}_{\rm in} \nonumber \\
  &=&  (-r_1+w_a^{\prime}t_1) \cos^2{\theta_{\rm pol}} \nonumber \\
  &\,& + (-r_1 e^{-i \Delta \phi_{\rm s_1}} + w_b^{\prime} t_1 e^{-i \frac{1}{2} \Delta \phi_{\rm t_1}}) \sin^2{\theta_{\rm pol}}  \label{E_refl_parallel}, \\
 \frac{E_{\rm refl \perp}}{E_{\rm in}} &=& M_{\rm refl} \vec{v}_{\rm in} \cdot R(\theta_{\rm pol})
 \begin{pmatrix}
0 \\
1 \\
\end{pmatrix}  \nonumber \\
  &=&  \left[ (-r_1+w_a^{\prime}t_1) - (-r_1e^{-i \Delta \phi_{\rm s_1}} + w_b^{\prime}t_1 e^{-i \frac{1}{2} \Delta \phi_{\rm t_1}}) \right] \nonumber \\
  &\,& \times \frac{\sin{(2\theta_{\rm pol})}}{2} \label{E_refl_orthogonal}.
\end{eqnarray}
These are similar to the electric fields of the bright reflection port and the dark antisymmetric port for a Fabry-P{\'e}rot-Michelson interferometer that has an unbalanced beam splitter.

The effects from the ETM birefringence can be calculated by setting $\Delta \phi_{\rm s_1}=\Delta \phi_{\rm t_1}=0$ and replacing $\Delta \phi_{\rm r_1}$ with $\Delta \phi_{\rm r_2}$ and $\theta_{\rm pol}$ with $\theta+\theta_{\rm pol}$. If we combine the effects from ITM and ETM, the phase of the reflected beam around the resonance can be approximated as
\begin{equation} \label{refl_phase}
\begin{split}
 \arg{\left( \frac{E_{\rm refl \parallel}}{E_{\rm in}} \right)} &= (\Delta \phi_{\rm s_1}-2 \Delta \phi_{\rm t_1})\sin^2{\theta_{\rm pol}} \\
 &-  \displaystyle{\frac{\mathcal{F}}{\pi}} \left[ \phi + \Delta \phi_{\rm r_1} \sin^2{\theta_{\rm pol}} \right. \\
 &+ \left. \Delta \phi_{\rm r_2} \sin^2{(\theta+\theta_{\rm pol})}  \right] ,
\end{split}
\end{equation}
with the approximation that $\Delta \phi_{{\rm r}_i} \ll 2 \pi/\mathcal{F}$ and $r_2=1$. It is clear that both the ETM rotation angle $\theta$ and the input beam polarization angle $\theta_{\rm pol}$ change the phase of the cavity reflected beam and will contribute to the phase noise, unless $\theta_{\rm pol}$ and $\theta+\theta_{\rm pol}$ are either 0 or $\pi/2$, where the effects are quadratic to these angles. The fluctuations of phase differences between ordinary and extraordinary axes also create phase noises, unless $\theta_{\rm pol}$ and $\theta+\theta_{\rm pol}$ are both 0.

It is worth noting that, even if we use this phase to lock the cavity, this does not generally mean that the cavity is locked on resonance to one of its polarization eigenmodes, as the cavity reflected beam contains the phase fluctuations from both polarization eigenmodes. To avoid the mixing of phase noises from two polarization eigenmodes, it is actually better to have higher static coating birefringence, i.e., $\Delta \phi_{{\rm r}_i} \gg 2 \pi/\mathcal{F}$. If the static coating birefringence is high such that one of the eigenmodes is out of resonance when the other is resonant, only $\Delta \phi_{\rm s_1}$ and $\phi$ terms remain in Eq.~(\ref{refl_phase}).

\section{Noises from birefringence} \label{Noises}
\begin{figure}[t]
\begin{center}
\includegraphics[width=\hsize]{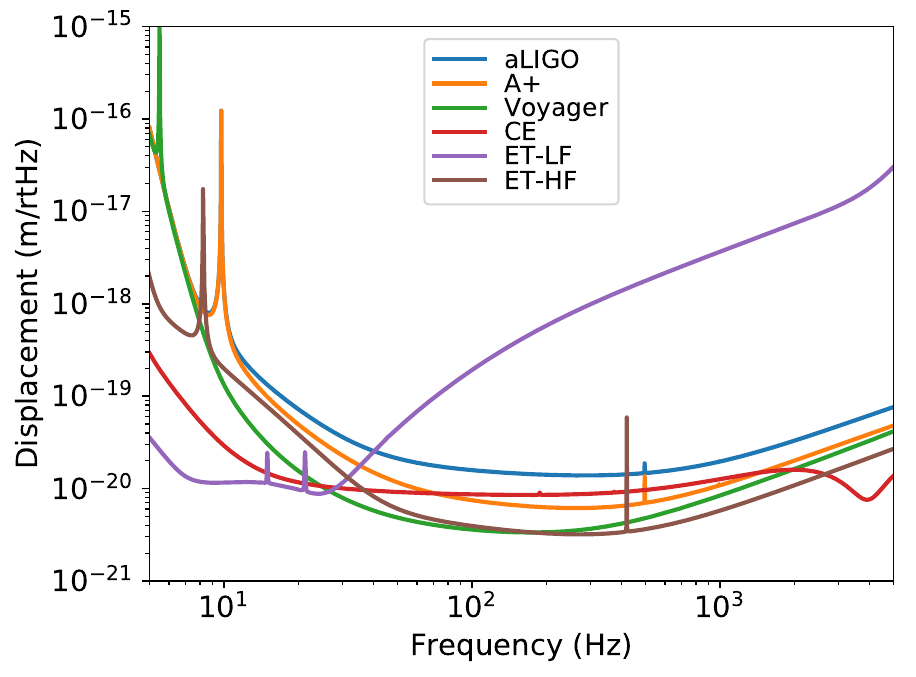}
\end{center}
\caption{\label{DisplacementSensitivity} The designed displacement sensitivity for different gravitational wave detectors. The strain sensitivity data are taken from Refs.~\cite{LIGO-T1500293,CE-T2000017,ET-0000A-18} and corrected to displacement sensitivities by removing frequency-dependent responses to gravitational waves~\cite{Essick2017}.}
\end{figure}

In this section, we calculate the phase noises from temporal fluctuations of birefringence and derive the requirements for the current and future gravitational wave detectors. For calculating the requirements, we have used the interferometer parameters summarized in Table~\ref{IFOParams} and the displacement sensitivity curves shown in Fig.~\ref{DisplacementSensitivity}. At the last part of this section, we also discuss the noise from the amplitude fluctuations in the orthogonal polarization at the antisymmetric port of the Fabry-P{\'e}rot Michelson interferometer. Although different interferometers plan to use different materials for the mirrors, discussions presented here do not depend on the choice of materials.

\begin{table}
    \caption{\label{IFOParams} Interferometer parameters of Advanced LIGO (aLIGO), A+, Voyager, Cosmic Explorer (CE), Einstein Telescope Low Frequency (ET-LF), and ET High Frequency (ET-HF) used for calculating requirements. $L$: arm length, $\mathcal{F}$: arm finesse, $t$: ITM thickness, $\lambda$: laser wavelength.}
\begin{ruledtabular}
\begin{tabular}{cccccc}
 & $L$ & $\mathcal{F}$ & $t$ & $\lambda$ & Ref. \\
\hline
aLIGO & 4~km & 450 & 20~cm & 1064~nm & \cite{aLIGO} \\
A+ & 4~km & 450 & 20~cm & 1064~nm & \cite{Aplus} \\
Voyager & 4~km & 3000 & 55~cm & 2050~nm & \cite{Voyager} \\
CE & 40~km & 450 & 27.3~cm & 2050~nm & \cite{CEHorizon} \\
ET-LF & 10~km & 900 & 57~cm & 1550~nm & \cite{ET-0007B-20} \\
ET-HF & 10~km & 900 & 30~cm & 1064~nm & \cite{ET-0007B-20} \\
\end{tabular}
\end{ruledtabular}
\end{table}

\subsection{Phase noises from substrate birefringence}
The phase changes from the ITM substrate birefringence can be calculated from Eq.~(\ref{refl_phase}) by setting $\Delta \phi_{\rm r_1} = \Delta \phi_{\rm r_2} = 0$ and $\Delta \phi_{\rm s_1} = \Delta \phi_{\rm t_1}$. In this case, Eq.~(\ref{refl_phase}) reduces to

\begin{equation} \label{refl_phase_substrate_only}
 \arg{\left( \frac{E_{\rm refl \parallel}}{E_{\rm in}} \right)} = -\Delta \phi_{\rm s_1} \sin^2{\theta_{\rm pol}} -  \displaystyle{\frac{\mathcal{F}}{\pi}} \phi .
\end{equation}
Therefore, the length noise couplings from the fluctuations of $\theta_{\rm pol}$ and $\Delta \phi_{\rm s_1}$ can be calculated as
\begin{eqnarray}
 \frac{\delta L}{\delta \theta_{\rm pol}} &=&  \frac{\lambda}{4 \pi} \frac{\delta [\arg{(E_{\rm refl \parallel})}]}{\delta \theta_{\rm pol}} \left( \frac{\delta [\arg{(E_{\rm refl \parallel})}]}{\delta \phi} \right)^{-1} \nonumber \\
 &=& \frac{\lambda}{4 \mathcal{F}} \Delta \phi_{\rm s_1} \sin{2\theta_{\rm pol}} ,\\
 \frac{\delta L}{\delta (\Delta \phi_{\rm s_1})} &=&  -\frac{\lambda}{4 \mathcal{F}} \sin^2{\theta_{\rm pol}} .
\end{eqnarray}

\subsection{Phase noises from coating birefringence}
Next, we consider the phase changes from the coating birefringence. From Eq.~(\ref{refl_phase}), it is clear that the second term from $\Delta \phi_{\rm r_1}$ and $\Delta \phi_{\rm r_2}$ contributes more to the phase of the reflected beam, compared with the first term from $\Delta \phi_{\rm s_1}$ and $\Delta \phi_{\rm t_1}$, since the phase acquired inside the cavity is enhanced by a factor of $\mathcal{F}/\pi$. The length noise couplings from the fluctuations of $\theta_{\rm pol}$, $\theta$, and $\Delta \phi_{{\rm r}_i}$ can be calculated as
\begin{eqnarray}
 \frac{\delta L}{\delta \theta_{\rm pol}} &=& \frac{\lambda}{4 \pi} [ \Delta \phi_{\rm r_1} \sin{2\theta_{\rm pol}} \nonumber \\
 &\,& + \Delta \phi_{\rm r_2} \sin{[2(\theta+\theta_{\rm pol})]} ],\\
 \frac{\delta L}{\delta \theta} &=& \frac{\lambda}{4 \pi} \Delta \phi_{\rm r_2} \sin{[2(\theta+\theta_{\rm pol})]} ,\\
 \frac{\delta L}{\delta (\Delta \phi_{\rm r_1})} &=&  -\frac{\lambda}{4 \pi} \sin^2{\theta_{\rm pol}} \label{coating_biref_req1} ,\\
 \frac{\delta L}{\delta (\Delta \phi_{\rm r_2})} &=&  -\frac{\lambda}{4 \pi} \sin^2{(\theta+\theta_{\rm pol})} \label{coating_biref_req2}.
\end{eqnarray}

\begin{figure}
\begin{center}
\includegraphics[width=\hsize]{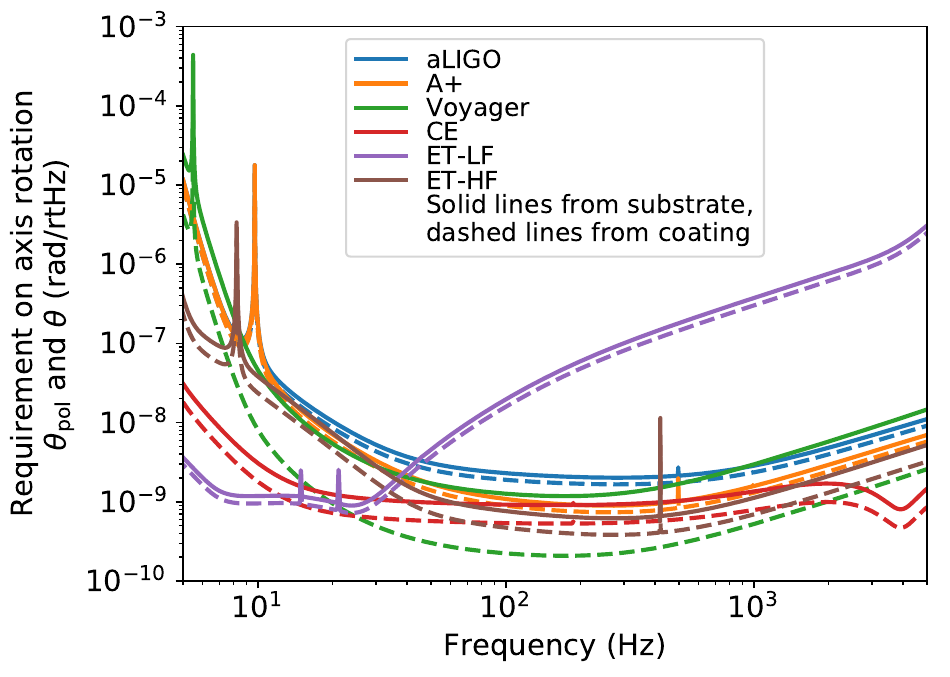}
\includegraphics[width=\hsize]{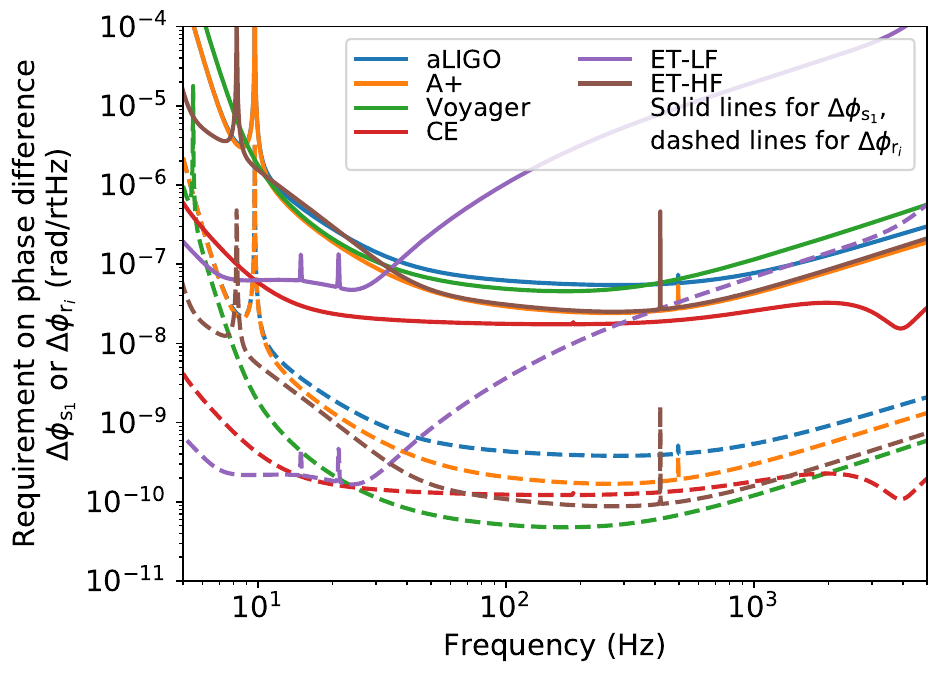}
\includegraphics[width=\hsize]{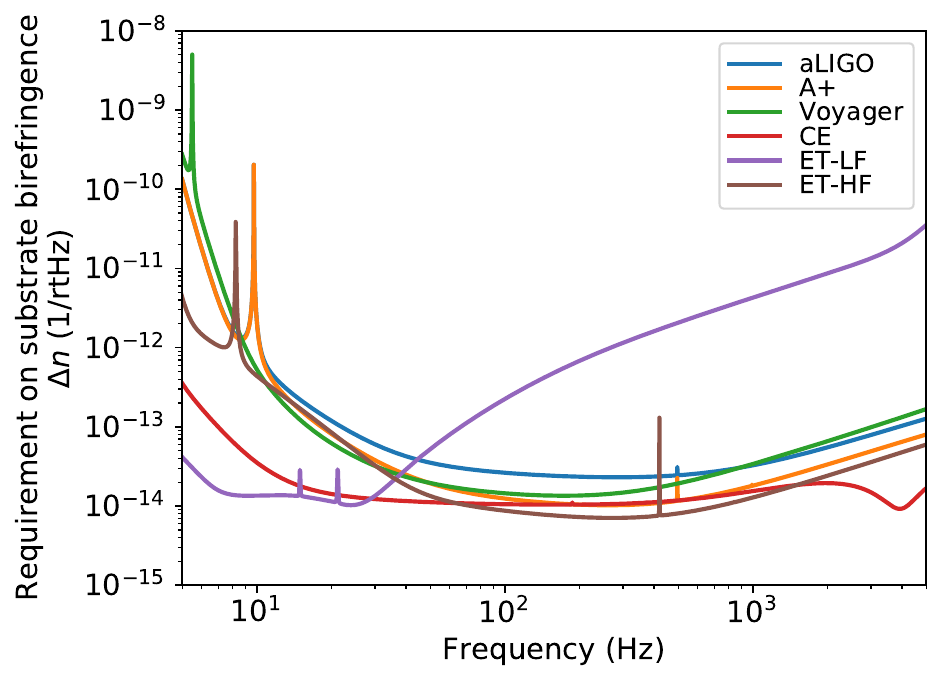}
\end{center}
\caption{\label{Requirements} The requirements on birefringence fluctuations from the axis rotations (top) and from the phase difference between ordinary and extraordinary axes (middle) for different gravitational wave detectors. The bottom plot shows the requirement on the substrate birefringence converted from the phase difference requirements on $\Delta \phi_{\rm s_1}$ in the middle plot, assuming uniform $\Delta n$, using Eq.~(\ref{uniform_thickness}). The solid lines are for the substrate that have a static birefringence of $\Delta n=10^{-7}$ and the dashed lines are for the coating that have a static birefringence of $\Delta \phi_{{\rm r}_i}=1$~mrad. For deriving these requirements, we assumed that the input beam polarization and the ETM axes are aligned to the ITM axes to $\theta_{\rm pol}=1^{\circ}$ and $\theta=1^{\circ}$, and no safety margin is considered.}
\end{figure}

\subsection{Requirements on birefringence fluctuations}
Noise couplings discussed above are nulled when $\theta_{\rm pol}=0$ and $\theta=0$. For KAGRA test masses, the sapphire $c$ axis was aligned to the cylindrical plane of the test mass within $0.1^{\circ}$~\cite{HiroseSapphire}. For deriving the requirements to birefringence fluctuations for the substrate and the coating, we assume that the input beam polarization and the ETM axes are aligned to the ITM axes to $\theta_{\rm pol}=1^{\circ}$ and $\theta=1^{\circ}$, respectively.

The solid lines in Fig.~\ref{Requirements} show the derived requirements for the substrate birefringence fluctuations. We assumed that the ITM substrate has uniform birefringence $\Delta n$, and $\Delta \phi_{\rm s_1}$ can be written using the mirror thickness $t$ as
\begin{equation} \label{uniform_thickness}
 \Delta \phi_{\rm s_1} = \frac{4 \pi}{\lambda} \Delta n t .
\end{equation}
We used the static birefringence value of $\Delta n=10^{-7}$, which is a typical measured value for silicon~\cite{AEISiliconBirefringence,UWASiliconBirefringence}. The dashed lines in Fig.~\ref{Requirements} show the derived requirements for the coating using the static birefringence value of $\Delta \phi_{{\rm r}_i}=1$~mrad, which is a typical measured value for AlGaAs coating~\cite{Cole2013,Winkler2021,Tanioka2023}. The requirements do not change for other materials when they have the same amount of static birefringence. For deriving the requirement for $\Delta \phi_{\rm r_j}$, we used Eq.~(\ref{coating_biref_req2}), as this gives more stringent requirement than Eq.~(\ref{coating_biref_req1}). All the requirements are divided by $\sqrt{2}$ to take into account of birefringence noises between two arm cavities to be incoherent, assuming both cavities have a similar level of birefringence. The requirements will be relaxed for common effects in two arms, such as the fluctuations in the input beam polarization angle and birefringence induced by laser intensity fluctuations.

The requirements on the axis rotations for future gravitational wave detectors is on the order of $10^{-10}$~rad/$\sqrt{\rm Hz}$. We note that the requirements on $\theta_{\rm pol}$ and $\theta$ presented here are also the requirements for the polarization fluctuation requirement for the input beam and the roll motion of the mirrors. As for the roll motion of the mirrors, the vertical seismic motion create less than $10^{-11}$~rad/$\sqrt{\rm Hz}$ level of roll motion above 10~Hz for the Advanced LIGO suspensions, if we conservatively assume that the coupling from vertical to roll motion is unity~\cite{ADAM-GD,aLIGOQuad}. Therefore, the birefringence noise from the roll motion of the mirrors is small enough.

The requirements on the phase differences between ordinary and extraordinary axes for future gravitational wave detectors are on the order of $10^{-8}$~rad/$\sqrt{\rm Hz}$ for the substrate, and $10^{-10}$~rad/$\sqrt{\rm Hz}$ for the coating. Birefringence at $10^{-8}$~rad/$\sqrt{\rm Hz}$ level can be feasibly evaluated with shot noise limited interferometry at the laser power of $P=10$~mW level, as the shot noise limited phase sensitivity of a Michelson interferometer is given by
\begin{equation}
 \phi_{\rm shot} = \sqrt{\frac{h c}{2 \lambda P}} ,
\end{equation}
where $h$ is Planck constant and $c$ is the speed of light. Evaluation of birefringence at $10^{-10}$~rad/$\sqrt{\rm Hz}$ level requires 10-W class laser or cavity enhancements. Measurements can be done at relatively lower power compared with gravitational wave detectors, as the phase noise from birefringence is attenuated by $\sin^2{\theta}$ and $\sin^2{(\theta+\theta_{\rm pol})}$, by aligning the polarization axis and the mirror crystal axes. In the evaluation setup, the phase noise can be enhanced by intentionally misaligning the axes.

One of the possible sources of birefringence fluctuations is magnetic field fluctuations due to Faraday effect. Measured magnetic field fluctuations at various gravitational wave detector sites are on the order of $10^{-12}$~T/$\sqrt{\rm Hz}$ at 10~Hz~\cite{Coughlin2018}, and the Verdet constant for silicon is 15~rad/(T$\cdot$m)~\cite{SiVerdet}. These give $10^{-11}$~rad/$\sqrt{\rm Hz}$ level of $\Delta \phi_{\rm s_1}$ for mirror thicknesses in Table~\ref{IFOParams}, which is below the requirements given above.

We note that, when deriving the requirements shown in Fig.~\ref{Requirements}, no safety margin was considered. This means that the designed sensitivity will be fully limited by one of the noises when that noise spectrum is the same as the requirement curve, and all the other noises are negligibly small. To achieve the design sensitivity, each noise should be negligibly smaller than the requirement, e.g., by a factor of 10.

\subsection{Amplitude noise at the antisymmetric port}
So far, we have considered the phase noise in the arm cavity reflected beams in gravitational wave detectors. In gravitational wave detectors, the differential arm length caused by gravitational waves will be read out as the interference fringe changes at the antisymmetric port. Birefringence fluctuations will also create power fluctuations in the orthogonal polarization, and it will be a noise source when the output Faraday isolator has a finite extinction ratio $\epsilon$, and the orthogonal polarization is not completely rejected. A slight misalignment of the axes between the input Faraday isolator and the output Faraday isolator would also cause a finite extinction ratio.

From Eq.~(\ref{E_refl_orthogonal}), the power of the cavity reflected beam in the orthogonal polarization from the birefringence in ITM can be written as
\begin{eqnarray}
 \frac{\left. P_{\rm refl \perp} \right|_{\rm res}}{P_{\rm in}} &\simeq& \frac{1}{4} \left( \Delta \phi_{\rm s_1} - 2\Delta \phi_{\rm t_1} - \frac{ \mathcal{F}}{\pi} \Delta \phi_{\rm r_1} \right)^2 \sin^2{(2 \theta_{\rm pol})} \nonumber , \\
 &\,& \label{power_loss_locked}
\end{eqnarray}
when the cavity is on resonance. Here, $P_{\rm in}=|E_{\rm in}|^2$ is the input power to the cavity, and we used that $r_2 = 1$, $r_1 \simeq 1$ and $t_1^2 = 1-r_1^2$, which are good approximations for arm cavities of gravitational wave detectors. We also assumed that the amount of birefringence is uniform and small, i.e., $\Delta \phi_{{\rm r}_i} \ll 2 \pi/\mathcal{F}$, $\Delta \phi_{\rm s_1} \ll 1$ and $\Delta \phi_{\rm t_1} \ll 1$.

As we can see from Eq.~(\ref{E_refl_orthogonal}), the orthogonal polarization is vanished when there is no birefringence, or $\theta_{\rm pol}$ is not 0 or $\pi/2$. The orthogonal polarization component is generated from the reflected electric field unbalance between two eigenmodes. Therefore, when the amount of birefringence is small, the phase of $E_{\rm refl \perp}$ is always around $\pi/2$ away from the phase of $E_{\rm refl \parallel}$. This means that the orthogonal polarization in the cavity reflection is always in the quadrature phase with respect to the gravitational wave signal, independent of the resonant condition of the cavity.

In the case of gravitational wave detectors, the antisymmetric port therefore will be at either the bright or the dark fringe for the orthogonal polarization, when it is at the dark fringe for the main polarization. When the both arms are completely symmetric and the amount of birefringence is the same, the antisymmetric port will be at the bright fringe for the orthogonal polarization. This is the same as the reason why the polarization signal from axion dark matter is present at the antisymmetric port, as discussed in Ref.~\cite{ADAM-GD}. In reality, the beam splitter in the Fabry-P{\'e}rot-Michelson interferometer adds extra phase difference between two polarization axes due to $\sim 45^{\circ}$ incident angle, and the fringe will be slightly shifted.

To derive the requirements for the extinction ratio $\epsilon$ of the output Faraday isolator, let us assume that the power of the orthogonal polarization component at the antisymmetric port can be roughly estimated from the power from one of the arms. By requiring the power fluctuation from the orthogonal polarization from one of the arms to be less than the shot noise of the local oscillator beam in the main polarization, we can require
\begin{equation}
 \epsilon < \frac{1}{\left. P_{\rm refl \perp} \right|_{\rm res}} \sqrt{\frac{2 h c P_{\rm LO}}{\lambda}} ,
\end{equation}
where $P_{\rm LO}$ is the power of the local oscillator beam at the antisymmetric port. When the requirements for the birefringence fluctuations derived in the previous subsections are met, the noise from the birefringence fluctuations are lower than the shot noise of the gravitational wave detector. Therefore, the requirement can be rewritten as
\begin{equation}
 \epsilon \lesssim \sqrt{\frac{P_{\rm LO}}{P_{\rm in}}} \left( \Delta \phi_{\rm s_1} - 2\Delta \phi_{\rm t_1} - \frac{ \mathcal{F}}{\pi} \Delta \phi_{\rm r_1} \right)^{-1} .
\end{equation}
For gravitational wave detectors operating with DC readout scheme~\cite{DCreadout}, $P_{\rm LO}$ and $P_{\rm in}$ are on the order of 10~mW and 10~kW for the power-recycled case, respectively. Assuming that the birefringence terms $\Delta \phi_{\rm s_1}$, $\Delta \phi_{\rm t_1}$, and $\Delta \phi_{\rm r_1} \mathcal{F}/\pi$ are on the order of 1~rad, the requirement to the extinction ratio will be $\epsilon \lesssim 0.1\%$. This means that the input Faraday isolator and the output Faraday isolator have to be aligned within $1.8^{\circ}$.

\section{Optical losses from inhomogeneous birefringence} \label{PowerLosses}
Birefringence and its inhomogeneity in cavities create power losses from depolarization. The mode content of the cavity reflected beam in the orthogonal polarization will be different depending on the locations of birefringence and the resonant condition of the cavity. In this section, we discuss the power of cavity reflected beam in the orthogonal polarization to estimate the optical loss.

To show that the different locations of birefringence create different mode content, we first consider the effects from ITM, as we have considered in Eqs.~(\ref{E_refl_parallel}) and (\ref{E_refl_orthogonal}). From Eq.~(\ref{E_refl_orthogonal}), the power losses to orthogonal polarization when the cavity is out of resonance will be
\begin{equation}
 \frac{\left. P_{\rm refl \perp} \right|_{\rm off}}{P_{\rm in}} \simeq \frac{1}{4} (\Delta \phi_{\rm s_1})^2 \sin^2{(2 \theta_{\rm pol})} \label{power_loss_unlocked}, \\
\end{equation}
under the same approximations used to derive Eq.~(\ref{power_loss_locked}).

So far, we have considered only the birefringence uniform over the substrate and the coating. When there is a perturbation from a uniform birefringence, spatial higher order modes are generated. The amount of the higher order modes in the orthogonal polarization can be estimated from inhomogeneous birefringence $\Delta \phi_{\rm s_1}^{\rm HOM}$. The power in the higher order modes when the cavity is on resonance and out of resonance will be
\begin{eqnarray}
 \frac{\left. P_{\rm refl \perp}^{\rm HOM} \right|_{\rm res}}{P_{\rm in}} &\simeq& \frac{1}{4} \left( \Delta \phi_{\rm s_1}^{\rm HOM} - \Delta \phi_{\rm t_1}^{\rm HOM} \right)^2 \sin^2{(2 \theta_{\rm pol})} \nonumber , \\
 &\,&  \label{power_loss_locked_HOM} \\
 \frac{\left. P_{\rm refl \perp}^{\rm HOM} \right|_{\rm off}}{P_{\rm in}} &\simeq& \frac{1}{4} (\Delta \phi_{\rm s_1}^{\rm HOM})^2 \sin^2{(2 \theta_{\rm pol})} \label{power_loss_unlocked_HOM},
\end{eqnarray}
respectively. Note that the coefficient for $\Delta \phi_{\rm t_1}^{\rm HOM}$ is 1, as opposed to 2 for $\Delta \phi_{\rm t_1}$ in Eq.~(\ref{power_loss_locked}), since higher order modes do not resonate in the cavity and higher order modes are generated in the ITM transmission of the intracavity beam.

For considering the effect from the ITM substrate birefringence, we can set $\Delta \phi_{\rm r_1}=0$, $\Delta \phi_{\rm s_1}=\Delta \phi_{\rm t_1}$ and $\Delta \phi_{\rm s_1}^{\rm HOM}=\Delta \phi_{\rm t_1}^{\rm HOM}$. In this case, the amount of the fundamental transverse mode in the orthogonal polarization stays the same when the cavity is out of resonance or on resonance. However, the amount of higher order modes in the orthogonal polarization is suppressed to the second order, as we can see from Eq.~(\ref{power_loss_locked_HOM}). This is similar to the Lawrence effect for the thermal lensing of ITM~\cite{Lawrence}. It is worth noting that the cavity reflected power in the main polarization $P_{\rm refl \parallel}$ could increase when the cavity is on resonance due to this effect, if the optical loss in the cavity is small compared with the optical loss from inhomogeneous birefringence.

For KAGRA sapphire ITM, the transmission wavefront error difference between two polarizations was measured to be around 60~nm in RMS~\cite{SapphireInhomogeneity,HiroseSapphire}, which corresponds to the round-trip phase difference $\Delta \phi_{\rm s_1}^{\rm HOM}$ of 0.7~rad in RMS. If we attribute this all to inhomogeneous refractive index difference using Eq.~(\ref{uniform_thickness}), this corresponds to $\Delta n^{\rm HOM}$ of $2 \times 10^{-7}$ in RMS, using the KAGRA sapphire mirror thickness being 15~cm and laser wavelength being 1064~nm. For sapphire, the amount of birefringence along the $c$ axis can be calculated with~\cite{Tokunari2006}
\begin{equation}
 \Delta n = \frac{n_o (n_o^2 - n_e^2) \psi^2}{n_e^2},
\end{equation}
where $n_e=1.747$ and $n_o=1.754$ are the refractive indices in the $c$ axis and in axes orthogonal to the $c$ axis, respectively, and $\psi \ll 1$ is the inclination of the light propagation direction with respect to the $c$ axis. Using this equation, the amount of birefringence observed in KAGRA can be explained by $\psi^{\rm HOM}$ being $0.2^{\circ}$ in RMS. This is larger than nominal orientation of the beam propagation axis with respect to the $c$ axis, which was aligned within $0.1^{\circ}$~\cite{HiroseSapphire}. This suggests that $\theta_{\rm pol}$ is also inhomogeneous and uncontrolled.

Using Eq.~(\ref{power_loss_unlocked_HOM}), this inhomogeneous birefringence create power loss to orthogonal polarization of around 10\% when the arm cavity is out of resonance, if $\theta_{\rm pol}$ is around $\pi/4$. This is consistent with the measured value in KAGRA, as reported in Ref.~\cite{PTEP01KAGRA}. The reduction of the power loss to orthogonal polarization on resonance was also observed, which is consistent with the Lawrence effect described above. In the KAGRA case, the power of the orthogonal polarization inside the power recycling cavity was reduced by a factor of 3 when the arm cavity was locked on resonance.

To make the optical loss due to inhomogeneous birefringence of ITM substrate always smaller than 0.1\%, $\Delta \phi_{\rm s_1}$ and $\Delta \phi_{\rm s_1}^{\rm HOM}$ need to be smaller than 0.06~rad in RMS. Achieving this with surface figuring alone could be challenging, as surface figuring cannot compensate for the phase difference between two axes. This requirement can be eased by aligning the input polarization axis to $\theta_{\rm pol}=0$ or $\pi/2$.

When considering the effect from the ITM coating birefringence, we can set $\Delta \phi_{\rm s_1}=\Delta \phi_{\rm r_1}$. However, $\Delta \phi_{\rm s_1}$ is not exactly $\Delta \phi_{\rm t_1}$, as the penetration length for the coating is different from the coating thickness. Therefore, the Lawrence effect does not completely suppress the higher order modes. If we can set $\Delta \phi_{\rm s_1}= l \Delta \phi_{\rm t_1}$, where $0<l<1$ is the ratio of the penetration length over the coating thickness, the higher order modes in the orthogonal polarization increase when the cavity is locked on resonance, for $l<0.5$. The fundamental transverse mode in the orthogonal polarization increases for high finesse cavities with $\mathcal{F}/\pi \gg 1$.

The mode content in the orthogonal polarization from the ETM coating birefringence can be obtained by replacing $\Delta \phi_{\rm r_1}$ with $\Delta \phi_{\rm r_2}$ and $\theta_{\rm pol}$ with $\theta+\theta_{\rm pol}$ in Eqs.~(\ref{power_loss_locked}), (\ref{power_loss_unlocked}), (\ref{power_loss_locked_HOM}), and (\ref{power_loss_unlocked_HOM}) and by setting $\Delta \phi_{\rm s_1} = \Delta \phi_{\rm t_1} = 0$, as
\begin{eqnarray}
 \frac{\left. P_{\rm refl \perp} \right|_{\rm res}}{P_{\rm in}} &\simeq& \frac{1}{4} \left(\frac{ \mathcal{F}}{\pi} \Delta \phi_{\rm r_2} \right)^2 \sin^2{[2(\theta+\theta_{\rm pol})]}, \label{power_loss_locked_ETM} \\
 \frac{\left. P_{\rm refl \perp} \right|_{\rm off}}{P_{\rm in}} &\simeq& 0, \\
 \frac{\left. P_{\rm refl \perp}^{\rm HOM} \right|_{\rm res}}{P_{\rm in}} &\simeq& 0, \\
 \frac{\left. P_{\rm refl \perp}^{\rm HOM} \right|_{\rm off}}{P_{\rm in}} &\simeq& 0.
\end{eqnarray}
Therefore, as for the effects from the ETM coating birefringence, the power in the orthogonal polarization increases when the cavity is locked on resonance, and the fundamental transverse mode dominates, because the higher order modes are suppressed in the cavity.

The discussions above highlights the fact that the optical losses from birefringence needs to be correctly taken into account to measure the optical losses in the arm cavity. It also suggests that, by measuring the mode content of the beam in the orthogonal polarization when the cavity is out of resonance and on resonance, we can estimate where the optical losses from birefringence are mainly coming from.

Future gravitational wave detector designs call for 10~dB of detected squeezing, requiring that the total optical loss be less than 10\%~\cite{Oelker2016}. From Eqs.~(\ref{power_loss_locked}) and (\ref{power_loss_locked_ETM}), $|\theta|$ and $|\theta+\theta_{\rm pol}|$ needs to be less than $1.8^{\circ}$, requiring the optical loss from birefringence be less than 0.1\%, when the birefringence terms $\Delta \phi_{\rm s_1}$, $\Delta \phi_{\rm t_1}$, and $\Delta \phi_{\rm r_j} \mathcal{F}/\pi$ are on the order of 1~rad. Similar to the discussions around Eq.~(\ref{optical_loss_reduced}), the polarization of the injected squeezed vacuum also needs to be aligned to less than $1.8^{\circ}$ to achieve the optical loss of less than 0.1\%.

\section{Conclusions and outlook} \label{Conclusion}
In this paper, we have discussed the effects of birefringence and its fluctuations in the mirror substrate and coating for laser interferometric gravitational wave detectors. We have shown that the polarization axis of the beam and the crystal axes of mirrors need to be aligned to minimize the optical losses and the noises from birefringence fluctuations. The optical losses from birefringence can be feasibly reduced to less than 0.1\%, when the axes are aligned within a few degrees. We have also shown that the requirements for the birefringence fluctuations in the substrate and the coating will be on the order of $10^{-8}$ and  $10^{-10}$ rad/$\sqrt{\rm Hz}$ at 100~Hz, respectively, for future gravitational wave detectors with mirrors that have $\Delta n=10^{-7}$ level of substrate birefringence and $\Delta \phi_{{\rm r}_i}=1$~mrad level of coating birefringence. When the static coating birefringence is large such that the resonant frequency difference between two polarization eigenmodes is larger than the cavity linewidth, the requirements on the coating birefringence fluctuations will be relaxed. In addition, we have derived the equations for estimating the amount of optical losses due to depolarization from inhomogeneous birefringence of mirror substrates and coatings. Our results provide the basic theory to study the noises and optical losses from birefringence fluctuations of mirrors in gravitational wave detectors.

In our model, we assumed that the amount of birefringence and misorientation of axes to be small. We also assumed that two interferometer arms of gravitational wave detectors to be close to symmetric. Detailed interferometer modeling will be necessary to treat larger birefringence, misorientation of axes, inhomogeneity of birefringence and axes orientations, and asymmetry between two arms including birefringent beam splitter effects. These effects would create classical radiation pressure noise, as intracavity power fluctuates from birefringence fluctuations. Including the power and signal recycling cavities to the model would also be important when these effects are not negligible and the resonant condition in the recycling cavities is different between polarizations. We leave these studies to future work.


\begin{acknowledgments}
We thank Hiroki Fujimoto, Kevin Kuns, Stefan W. Ballmer, Valery Frolov and Martin M. Fejer for insightful discussions. This work was supported by the Gordon and Betty Moore Foundation, by the National Science Foundation under Grant No.~PHY-1912677, by JSPS KAKENHI Grant No.~JP20H05854, and by JST PRESTO Grant No.~JPMJPR200B. F.S.--C. acknowledges support from the Barish--Weiss postdoctoral fellowship.

This paper carries LIGO DCC No.~LIGO-P2300220 and JGW Document No.~JGW-P2315068. 
\end{acknowledgments}

\appendix
\section{Derivation of electric fields} \label{CavityDerivation}
Here we derive the electric field inside the cavity in Eqs.~\eqref{eq:Ecav} and \eqref{eq:Ecavprime}, and the electric field of the cavity reflection in Eq.~\eqref{eq:Erefl}.

In the basis of ITM crystal axes, the amplitude reflectivity of ETM can be written as $R(-\theta) R_2 R(\theta)$~\cite{JonesCalculus}. The rotation matrix $R(\theta)$ is necessary to take into account of the axes rotation between ITM and ETM. Therefore, the Jones matrix for the cavity round-trip can be written as a product of ITM reflection, ETM reflection, and the phase shift accumulated in the round-trip as
\begin{equation}
 A = R_1 R(-\theta) R_2 R(\theta) e^{-i \phi} .
\end{equation}
The electric field inside the cavity that propagates from ITM to ETM is a sum of the ITM transmitted field and its multiple reflections inside the cavity, which can be written as
\begin{eqnarray}
 \vec{E}_{\rm cav} &=& T_1 \vec{E}_{\rm in} + A T_1 \vec{E}_{\rm in} + A^2 T_1 \vec{E}_{\rm in} + \cdots \\
 &=& \sum_{n=1}^{\infty} A^{n-1} T_1 \vec{E}_{\rm in} .
\end{eqnarray}
This is a sum of an infinite geometric series, and Eq.~\eqref{eq:Ecav} can be derived.

The electric field inside the cavity that propagates from ETM to ITM has an additional reflection from ETM and phase $\phi$ from a cavity round-trip, which lead to Eq.~\eqref{eq:Ecavprime}. The electric field of the cavity reflection is the sum of the field reflected from ITM substrate side and the intracavity field transmitted through ITM. Therefore, it can be written as
\begin{equation}
 \vec{E}_{\rm refl} = S_1 \vec{E}_{\rm in} + T_1 \vec{E}^{\prime}_{\rm cav} ,
\end{equation}
and Eq.~\eqref{eq:Erefl} can be derived.

\section{Derivation of equivalent phase anisotropy} \label{EqPhaseDerivation}
Here we derive the equivalent phase anisotropy in Eq.~\eqref{eq:deltaEQ}. We consider the situation described in Ref.~\cite{Brandi1997}, where the phase anisotropy and relative orientation of the birefringent cavity are captured by a single equivalent Jones transformation. To simplify the notation, we write the Jones operators in the Pauli basis spanned by $I$ and $\vec{\sigma}$ where $I$ is the identity matrix and $\vec\sigma = \sigma_{\rm o} \vec{e}_{\rm o} + \sigma_{\rm e} \vec{e}_{\rm e} + \sigma_{\rm z} \vec{e}_{\rm z} $ is the Pauli vector used to map rotations along the ordinary, extraordinary, and cavity axis unit vectors, respectively. For example, the Jones operator for a half-wave plate with phase anisotropy $\delta$ oriented at an angle $\theta$ away from the ordinary axis $\vec{e}_{\rm o}$ may be written in this representation as
\begin{equation}
    \vec{\mathcal W}(\delta, \theta)\cdot \vec \sigma = \cos\left(\frac{\delta}{2}\right) {I} - i \sin\left(\frac{\delta}{2}\right) \left(\sin2\theta {\sigma}_{\rm o} -\cos2\theta {\sigma}_{\rm z}\right)
\end{equation}
and reduced to $\cos\left(\frac{\delta}{2}\right) I + i \sin\left(\frac{\delta}{2}\right) {\sigma}_{\rm z} $ when aligned with $\vec{e}_{\rm o}$ (i.e. $\theta=0$). 

Following Ref.~\cite{Brandi1997}, the equivalent wave plate anisotropy $\delta_{\rm EQ}$ comprises two wave plate operators with phase anisotropies $\delta_1$ and $\delta_2$ oriented at $\theta_1 = 0$ and $\theta_2 = \theta_{\rm WP}$ respectively. Then, the total operator
\begin{equation}
    \vec{\mathcal W}(\delta_{\rm EQ}, \theta_{\rm EQ}) \cdot \vec \sigma = [\vec{\mathcal W}(\delta_1, 0)  \cdot \vec \sigma][\vec{\mathcal W}(\delta_2, \theta_{\rm WP}) \cdot \vec \sigma]
\end{equation}
can be constructed by the individual operators. After some manipulation, we obtain an equation for each component beginning with
\begin{align}\label{eq:app1}
    \cos\left(\frac{\delta_{\rm EQ}}{2}\right) &= \cos\left(\frac{\delta_1}{2}\right)\cos\left(\frac{\delta_2}{2}\right) \\\nonumber
    &  - \sin\left(\frac{\delta_1}{2}\right) \sin \left(\frac{\delta_2}{2} \right) \cos2\theta_{\rm WP}
\end{align}
from terms along $I$, and then
\begin{align}\label{eq:app2}
    \sin2\theta_{\rm EQ} \sin \left(\frac{\delta_{\rm EQ}}{2}\right) =
    \cos\left(\frac{\delta_1}{2}\right) \sin \left(\frac{\delta_2}{2}\right) \sin2\theta_{\rm WP}
\end{align}
for terms along $\sigma_{\rm o}$, 
\begin{align}\label{eq:app3}
    \sin\left(\frac{\delta_1}{2}\right) \sin \left(\frac{\delta_2}{2}\right) \sin2\theta_{\rm WP} = 0
\end{align}
from terms along $\sigma_{\rm e}$, and
\begin{align}\label{eq:app4}
    \sin\left(\frac{\delta_{\rm EQ}}{2}\right)\cos2\theta_{\rm EQ} &= \sin\left(\frac{\delta_1}{2}\right)\cos\left(\frac{\delta_2}{2}\right)\\\nonumber
    & +\cos\left(\frac{\delta_1}{2}\right)\sin\left(\frac{\delta_2}{2}\right)\cos2\theta_{\rm WP}
\end{align}
for terms along $\sigma_{\rm z}$. 

Letting $\delta_{\rm EQ} \ll 1$, $\delta_1 \ll 1$, and $\delta_2 \ll1$, we may expand Eqs~\eqref{eq:app1}--\eqref{eq:app4} to second order in $\delta_{\rm EQ}$, $\delta_1$, and $\delta_2$, keeping terms only of up to ${\mathcal O}(\delta^2)$. Then, Eq.~\eqref{eq:app1} becomes
\begin{equation}
	1 - \frac{\delta_{\rm EQ}^2}{8} \approx 1 - \frac{\delta_1^2}{8} - \frac{\delta_2^2}{8} - \frac{\delta_1 \delta_2}{4} \cos2\theta_{\rm WP}.
\end{equation}
from which Eq.~\eqref{eq:deltaEQ} may be easily derived and applied for the cases discussed in the text when $\delta_i = \Delta \phi_{r_i}$. Finally, inserting~\eqref{eq:deltaEQ} back into~\eqref{eq:app4} gives
\begin{align}
   \cos2\theta_{\rm EQ} &\approx \frac{\delta_1+\delta_2\cos2\theta_{\rm WP}}{\sqrt{\delta_1^2 - \delta_2^2 + 4 \delta_1\delta_2 \cos2\theta_{\rm WP}}}.
\end{align}
from which Eq.~\eqref{thetaEQ} may be easily derived and applied for the cases described in text when $\delta_i = \Delta \phi_{r_i}$.

\bibliography{birefringence}

\begin{thebibliography}{50}%
\makeatletter
\providecommand \@ifxundefined [1]{%
 \@ifx{#1\undefined}
}%
\providecommand \@ifnum [1]{%
 \ifnum #1\expandafter \@firstoftwo
 \else \expandafter \@secondoftwo
 \fi
}%
\providecommand \@ifx [1]{%
 \ifx #1\expandafter \@firstoftwo
 \else \expandafter \@secondoftwo
 \fi
}%
\providecommand \natexlab [1]{#1}%
\providecommand \enquote  [1]{``#1''}%
\providecommand \bibnamefont  [1]{#1}%
\providecommand \bibfnamefont [1]{#1}%
\providecommand \citenamefont [1]{#1}%
\providecommand \href@noop [0]{\@secondoftwo}%
\providecommand \href [0]{\begingroup \@sanitize@url \@href}%
\providecommand \@href[1]{\@@startlink{#1}\@@href}%
\providecommand \@@href[1]{\endgroup#1\@@endlink}%
\providecommand \@sanitize@url [0]{\catcode `\\12\catcode `\$12\catcode
  `\&12\catcode `\#12\catcode `\^12\catcode `\_12\catcode `\%12\relax}%
\providecommand \@@startlink[1]{}%
\providecommand \@@endlink[0]{}%
\providecommand \url  [0]{\begingroup\@sanitize@url \@url }%
\providecommand \@url [1]{\endgroup\@href {#1}{\urlprefix }}%
\providecommand \urlprefix  [0]{URL }%
\providecommand \Eprint [0]{\href }%
\providecommand \doibase [0]{https://doi.org/}%
\providecommand \selectlanguage [0]{\@gobble}%
\providecommand \bibinfo  [0]{\@secondoftwo}%
\providecommand \bibfield  [0]{\@secondoftwo}%
\providecommand \translation [1]{[#1]}%
\providecommand \BibitemOpen [0]{}%
\providecommand \bibitemStop [0]{}%
\providecommand \bibitemNoStop [0]{.\EOS\space}%
\providecommand \EOS [0]{\spacefactor3000\relax}%
\providecommand \BibitemShut  [1]{\csname bibitem#1\endcsname}%
\let\auto@bib@innerbib\@empty
\bibitem [{\citenamefont {Abbott}\ \emph {et~al.}(2016)\citenamefont {Abbott}
  \emph {et~al.}}]{GW150914}%
  \BibitemOpen
  \bibfield  {author} {\bibinfo {author} {\bibfnamefont {B.~P.}\ \bibnamefont
  {Abbott}} \emph {et~al.} (\bibinfo {collaboration} {LIGO Scientific
  Collaboration and Virgo Collaboration}),\ }\href
  {https://doi.org/10.1103/PhysRevLett.116.061102} {\bibfield  {journal}
  {\bibinfo  {journal} {Phys. Rev. Lett.}\ }\textbf {\bibinfo {volume} {116}},\
  \bibinfo {pages} {061102} (\bibinfo {year} {2016})}\BibitemShut {NoStop}%
\bibitem [{\citenamefont {Abbott}\ \emph
  {et~al.}(2017{\natexlab{a}})\citenamefont {Abbott} \emph
  {et~al.}}]{GW170817}%
  \BibitemOpen
  \bibfield  {author} {\bibinfo {author} {\bibfnamefont {B.~P.}\ \bibnamefont
  {Abbott}} \emph {et~al.} (\bibinfo {collaboration} {LIGO Scientific
  Collaboration and Virgo Collaboration}),\ }\href
  {https://doi.org/10.1103/PhysRevLett.119.161101} {\bibfield  {journal}
  {\bibinfo  {journal} {Phys. Rev. Lett.}\ }\textbf {\bibinfo {volume} {119}},\
  \bibinfo {pages} {161101} (\bibinfo {year} {2017}{\natexlab{a}})}\BibitemShut
  {NoStop}%
\bibitem [{\citenamefont {Abbott}\ \emph
  {et~al.}(2017{\natexlab{b}})\citenamefont {Abbott} \emph
  {et~al.}}]{Multimessenger}%
  \BibitemOpen
  \bibfield  {author} {\bibinfo {author} {\bibfnamefont {B.~P.}\ \bibnamefont
  {Abbott}} \emph {et~al.},\ }\href
  {http://stacks.iop.org/2041-8205/848/i=2/a=L12} {\bibfield  {journal}
  {\bibinfo  {journal} {The Astrophysical Journal Letters}\ }\textbf {\bibinfo
  {volume} {848}},\ \bibinfo {pages} {L12} (\bibinfo {year}
  {2017}{\natexlab{b}})}\BibitemShut {NoStop}%
\bibitem [{\citenamefont {Aasi}\ \emph {et~al.}(2015)\citenamefont {Aasi} \emph
  {et~al.}}]{aLIGO}%
  \BibitemOpen
  \bibfield  {author} {\bibinfo {author} {\bibfnamefont {J.}~\bibnamefont
  {Aasi}} \emph {et~al.} (\bibinfo {collaboration} {The LIGO Scientific
  Collaboration}),\ }\href {http://stacks.iop.org/0264-9381/32/i=7/a=074001}
  {\bibfield  {journal} {\bibinfo  {journal} {Classical and Quantum Gravity}\
  }\textbf {\bibinfo {volume} {32}},\ \bibinfo {pages} {074001} (\bibinfo
  {year} {2015})}\BibitemShut {NoStop}%
\bibitem [{\citenamefont {Acernese}\ \emph {et~al.}(2015)\citenamefont
  {Acernese} \emph {et~al.}}]{AdV}%
  \BibitemOpen
  \bibfield  {author} {\bibinfo {author} {\bibfnamefont {F.}~\bibnamefont
  {Acernese}} \emph {et~al.} (\bibinfo {collaboration} {Virgo Collaboration}),\
  }\href {http://stacks.iop.org/0264-9381/32/i=2/a=024001} {\bibfield
  {journal} {\bibinfo  {journal} {Classical and Quantum Gravity}\ }\textbf
  {\bibinfo {volume} {32}},\ \bibinfo {pages} {024001} (\bibinfo {year}
  {2015})}\BibitemShut {NoStop}%
\bibitem [{\citenamefont {Abbott}\ \emph {et~al.}(2020)\citenamefont {Abbott}
  \emph {et~al.}}]{ObservingScenarioPaper}%
  \BibitemOpen
  \bibfield  {author} {\bibinfo {author} {\bibfnamefont {B.~P.}\ \bibnamefont
  {Abbott}} \emph {et~al.},\ }\href
  {https://doi.org/10.1007/s41114-020-00026-9} {\bibfield  {journal} {\bibinfo
  {journal} {Living Reviews in Relativity}\ }\textbf {\bibinfo {volume} {23}},\
  \bibinfo {pages} {3} (\bibinfo {year} {2020})}\BibitemShut {NoStop}%
\bibitem [{\citenamefont {Adhikari}(2014)}]{RanaRMP}%
  \BibitemOpen
  \bibfield  {author} {\bibinfo {author} {\bibfnamefont {R.~X.}\ \bibnamefont
  {Adhikari}},\ }\href {https://doi.org/10.1103/RevModPhys.86.121} {\bibfield
  {journal} {\bibinfo  {journal} {Rev. Mod. Phys.}\ }\textbf {\bibinfo {volume}
  {86}},\ \bibinfo {pages} {121} (\bibinfo {year} {2014})}\BibitemShut
  {NoStop}%
\bibitem [{\citenamefont {Aso}\ \emph {et~al.}(2013)\citenamefont {Aso},
  \citenamefont {Michimura}, \citenamefont {Somiya}, \citenamefont {Ando},
  \citenamefont {Miyakawa}, \citenamefont {Sekiguchi}, \citenamefont
  {Tatsumi},\ and\ \citenamefont {Yamamoto}}]{AsoKAGRA}%
  \BibitemOpen
  \bibfield  {author} {\bibinfo {author} {\bibfnamefont {Y.}~\bibnamefont
  {Aso}}, \bibinfo {author} {\bibfnamefont {Y.}~\bibnamefont {Michimura}},
  \bibinfo {author} {\bibfnamefont {K.}~\bibnamefont {Somiya}}, \bibinfo
  {author} {\bibfnamefont {M.}~\bibnamefont {Ando}}, \bibinfo {author}
  {\bibfnamefont {O.}~\bibnamefont {Miyakawa}}, \bibinfo {author}
  {\bibfnamefont {T.}~\bibnamefont {Sekiguchi}}, \bibinfo {author}
  {\bibfnamefont {D.}~\bibnamefont {Tatsumi}},\ and\ \bibinfo {author}
  {\bibfnamefont {H.}~\bibnamefont {Yamamoto}} (\bibinfo {collaboration} {The
  KAGRA Collaboration}),\ }\href {https://doi.org/10.1103/PhysRevD.88.043007}
  {\bibfield  {journal} {\bibinfo  {journal} {Phys. Rev. D}\ }\textbf {\bibinfo
  {volume} {88}},\ \bibinfo {pages} {043007} (\bibinfo {year}
  {2013})}\BibitemShut {NoStop}%
\bibitem [{\citenamefont {Akutsu}\ \emph {et~al.}(2021)\citenamefont {Akutsu}
  \emph {et~al.}}]{PTEP01KAGRA}%
  \BibitemOpen
  \bibfield  {author} {\bibinfo {author} {\bibfnamefont {T.}~\bibnamefont
  {Akutsu}} \emph {et~al.} (\bibinfo {collaboration} {The KAGRA
  Collaboration}),\ }\href {https://doi.org/10.1093/ptep/ptaa125} {\bibfield
  {journal} {\bibinfo  {journal} {Progress of Theoretical and Experimental
  Physics}\ }\textbf {\bibinfo {volume} {2021}},\ \bibinfo {pages} {05A101}
  (\bibinfo {year} {2021})}\BibitemShut {NoStop}%
\bibitem [{\citenamefont {Michimura}\ \emph {et~al.}(2018)\citenamefont
  {Michimura}, \citenamefont {Komori}, \citenamefont {Nishizawa}, \citenamefont
  {Takeda}, \citenamefont {Nagano}, \citenamefont {Enomoto}, \citenamefont
  {Hayama}, \citenamefont {Somiya},\ and\ \citenamefont {Ando}}]{PSOKAGRA}%
  \BibitemOpen
  \bibfield  {author} {\bibinfo {author} {\bibfnamefont {Y.}~\bibnamefont
  {Michimura}}, \bibinfo {author} {\bibfnamefont {K.}~\bibnamefont {Komori}},
  \bibinfo {author} {\bibfnamefont {A.}~\bibnamefont {Nishizawa}}, \bibinfo
  {author} {\bibfnamefont {H.}~\bibnamefont {Takeda}}, \bibinfo {author}
  {\bibfnamefont {K.}~\bibnamefont {Nagano}}, \bibinfo {author} {\bibfnamefont
  {Y.}~\bibnamefont {Enomoto}}, \bibinfo {author} {\bibfnamefont
  {K.}~\bibnamefont {Hayama}}, \bibinfo {author} {\bibfnamefont
  {K.}~\bibnamefont {Somiya}},\ and\ \bibinfo {author} {\bibfnamefont
  {M.}~\bibnamefont {Ando}},\ }\href
  {https://doi.org/10.1103/PhysRevD.97.122003} {\bibfield  {journal} {\bibinfo
  {journal} {Phys. Rev. D}\ }\textbf {\bibinfo {volume} {97}},\ \bibinfo
  {pages} {122003} (\bibinfo {year} {2018})}\BibitemShut {NoStop}%
\bibitem [{\citenamefont {Adhikari}\ \emph {et~al.}(2020)\citenamefont
  {Adhikari} \emph {et~al.}}]{Voyager}%
  \BibitemOpen
  \bibfield  {author} {\bibinfo {author} {\bibfnamefont {R.~X.}\ \bibnamefont
  {Adhikari}} \emph {et~al.},\ }\href
  {https://doi.org/10.1088/1361-6382/ab9143} {\bibfield  {journal} {\bibinfo
  {journal} {Classical and Quantum Gravity}\ }\textbf {\bibinfo {volume}
  {37}},\ \bibinfo {pages} {165003} (\bibinfo {year} {2020})}\BibitemShut
  {NoStop}%
\bibitem [{\citenamefont {Punturo}\ \emph {et~al.}(2010)\citenamefont {Punturo}
  \emph {et~al.}}]{ET}%
  \BibitemOpen
  \bibfield  {author} {\bibinfo {author} {\bibfnamefont {M.}~\bibnamefont
  {Punturo}} \emph {et~al.},\ }\href
  {http://stacks.iop.org/0264-9381/27/i=19/a=194002} {\bibfield  {journal}
  {\bibinfo  {journal} {Classical and Quantum Gravity}\ }\textbf {\bibinfo
  {volume} {27}},\ \bibinfo {pages} {194002} (\bibinfo {year}
  {2010})}\BibitemShut {NoStop}%
\bibitem [{\citenamefont {{ET~steering~committee}}(2020)}]{ET-0007B-20}%
  \BibitemOpen
  \bibfield  {author} {\bibinfo {author} {\bibnamefont
  {{ET~steering~committee}}},\ }\href {https://apps.et-gw.eu/tds/ql/?c=15418}
  {\emph {\bibinfo {title} {ET design report update 2020}}},\ \bibinfo {type}
  {Tech. Rep.}\ \bibinfo {number} {ET-0007B-20}\ (\bibinfo {year}
  {2020})\BibitemShut {NoStop}%
\bibitem [{\citenamefont {Abbott}\ \emph
  {et~al.}(2017{\natexlab{c}})\citenamefont {Abbott} \emph {et~al.}}]{CE}%
  \BibitemOpen
  \bibfield  {author} {\bibinfo {author} {\bibfnamefont {B.~P.}\ \bibnamefont
  {Abbott}} \emph {et~al.} (\bibinfo {collaboration} {LIGO Scientific
  Collaboration}),\ }\href {http://stacks.iop.org/0264-9381/34/i=4/a=044001}
  {\bibfield  {journal} {\bibinfo  {journal} {Classical and Quantum Gravity}\
  }\textbf {\bibinfo {volume} {34}},\ \bibinfo {pages} {044001} (\bibinfo
  {year} {2017}{\natexlab{c}})}\BibitemShut {NoStop}%
\bibitem [{\citenamefont {Evans}\ \emph {et~al.}(2021)\citenamefont {Evans},
  \citenamefont {Adhikari}, \citenamefont {Afle}, \citenamefont {Ballmer},
  \citenamefont {Biscoveanu}, \citenamefont {Borhanian}, \citenamefont {Brown},
  \citenamefont {Chen}, \citenamefont {Eisenstein}, \citenamefont {Gruson},
  \citenamefont {Gupta}, \citenamefont {Hall}, \citenamefont {Huxford},
  \citenamefont {Kamai}, \citenamefont {Kashyap}, \citenamefont {Kissel},
  \citenamefont {Kuns}, \citenamefont {Landry}, \citenamefont {Lenon},
  \citenamefont {Lovelace}, \citenamefont {McCuller}, \citenamefont {Ng},
  \citenamefont {Nitz}, \citenamefont {Read}, \citenamefont {Sathyaprakash},
  \citenamefont {Shoemaker}, \citenamefont {Slagmolen}, \citenamefont {Smith},
  \citenamefont {Srivastava}, \citenamefont {Sun}, \citenamefont {Vitale},\
  and\ \citenamefont {Weiss}}]{CEHorizon}%
  \BibitemOpen
  \bibfield  {author} {\bibinfo {author} {\bibfnamefont {M.}~\bibnamefont
  {Evans}}, \bibinfo {author} {\bibfnamefont {R.~X.}\ \bibnamefont {Adhikari}},
  \bibinfo {author} {\bibfnamefont {C.}~\bibnamefont {Afle}}, \bibinfo {author}
  {\bibfnamefont {S.~W.}\ \bibnamefont {Ballmer}}, \bibinfo {author}
  {\bibfnamefont {S.}~\bibnamefont {Biscoveanu}}, \bibinfo {author}
  {\bibfnamefont {S.}~\bibnamefont {Borhanian}}, \bibinfo {author}
  {\bibfnamefont {D.~A.}\ \bibnamefont {Brown}}, \bibinfo {author}
  {\bibfnamefont {Y.}~\bibnamefont {Chen}}, \bibinfo {author} {\bibfnamefont
  {R.}~\bibnamefont {Eisenstein}}, \bibinfo {author} {\bibfnamefont
  {A.}~\bibnamefont {Gruson}}, \bibinfo {author} {\bibfnamefont
  {A.}~\bibnamefont {Gupta}}, \bibinfo {author} {\bibfnamefont {E.~D.}\
  \bibnamefont {Hall}}, \bibinfo {author} {\bibfnamefont {R.}~\bibnamefont
  {Huxford}}, \bibinfo {author} {\bibfnamefont {B.}~\bibnamefont {Kamai}},
  \bibinfo {author} {\bibfnamefont {R.}~\bibnamefont {Kashyap}}, \bibinfo
  {author} {\bibfnamefont {J.~S.}\ \bibnamefont {Kissel}}, \bibinfo {author}
  {\bibfnamefont {K.}~\bibnamefont {Kuns}}, \bibinfo {author} {\bibfnamefont
  {P.}~\bibnamefont {Landry}}, \bibinfo {author} {\bibfnamefont
  {A.}~\bibnamefont {Lenon}}, \bibinfo {author} {\bibfnamefont
  {G.}~\bibnamefont {Lovelace}}, \bibinfo {author} {\bibfnamefont
  {L.}~\bibnamefont {McCuller}}, \bibinfo {author} {\bibfnamefont {K.~K.~Y.}\
  \bibnamefont {Ng}}, \bibinfo {author} {\bibfnamefont {A.~H.}\ \bibnamefont
  {Nitz}}, \bibinfo {author} {\bibfnamefont {J.}~\bibnamefont {Read}}, \bibinfo
  {author} {\bibfnamefont {B.~S.}\ \bibnamefont {Sathyaprakash}}, \bibinfo
  {author} {\bibfnamefont {D.~H.}\ \bibnamefont {Shoemaker}}, \bibinfo {author}
  {\bibfnamefont {B.~J.~J.}\ \bibnamefont {Slagmolen}}, \bibinfo {author}
  {\bibfnamefont {J.~R.}\ \bibnamefont {Smith}}, \bibinfo {author}
  {\bibfnamefont {V.}~\bibnamefont {Srivastava}}, \bibinfo {author}
  {\bibfnamefont {L.}~\bibnamefont {Sun}}, \bibinfo {author} {\bibfnamefont
  {S.}~\bibnamefont {Vitale}},\ and\ \bibinfo {author} {\bibfnamefont
  {R.}~\bibnamefont {Weiss}},\ }\href@noop {} {\bibinfo {title} {A horizon
  study for cosmic explorer: Science, observatories, and community}} (\bibinfo
  {year} {2021}),\ \Eprint {https://arxiv.org/abs/2109.09882} {arXiv:2109.09882
  [astro-ph.IM]} \BibitemShut {NoStop}%
\bibitem [{\citenamefont {Cole}\ \emph {et~al.}(2013)\citenamefont {Cole},
  \citenamefont {Zhang}, \citenamefont {Martin}, \citenamefont {Ye},\ and\
  \citenamefont {Aspelmeyer}}]{Cole2013}%
  \BibitemOpen
  \bibfield  {author} {\bibinfo {author} {\bibfnamefont {G.~D.}\ \bibnamefont
  {Cole}}, \bibinfo {author} {\bibfnamefont {W.}~\bibnamefont {Zhang}},
  \bibinfo {author} {\bibfnamefont {M.~J.}\ \bibnamefont {Martin}}, \bibinfo
  {author} {\bibfnamefont {J.}~\bibnamefont {Ye}},\ and\ \bibinfo {author}
  {\bibfnamefont {M.}~\bibnamefont {Aspelmeyer}},\ }\href
  {https://doi.org/10.1038/nphoton.2013.174} {\bibfield  {journal} {\bibinfo
  {journal} {Nature Photonics}\ }\textbf {\bibinfo {volume} {7}},\ \bibinfo
  {pages} {644} (\bibinfo {year} {2013})}\BibitemShut {NoStop}%
\bibitem [{\citenamefont {Cumming}\ \emph {et~al.}(2015)\citenamefont
  {Cumming}, \citenamefont {Craig}, \citenamefont {Martin}, \citenamefont
  {Bassiri}, \citenamefont {Cunningham}, \citenamefont {Fejer}, \citenamefont
  {Harris}, \citenamefont {Haughian}, \citenamefont {Heinert}, \citenamefont
  {Lantz}, \citenamefont {Lin}, \citenamefont {Markosyan}, \citenamefont
  {Nawrodt}, \citenamefont {Route},\ and\ \citenamefont {Rowan}}]{AlGaP}%
  \BibitemOpen
  \bibfield  {author} {\bibinfo {author} {\bibfnamefont {A.~V.}\ \bibnamefont
  {Cumming}}, \bibinfo {author} {\bibfnamefont {K.}~\bibnamefont {Craig}},
  \bibinfo {author} {\bibfnamefont {I.~W.}\ \bibnamefont {Martin}}, \bibinfo
  {author} {\bibfnamefont {R.}~\bibnamefont {Bassiri}}, \bibinfo {author}
  {\bibfnamefont {L.}~\bibnamefont {Cunningham}}, \bibinfo {author}
  {\bibfnamefont {M.~M.}\ \bibnamefont {Fejer}}, \bibinfo {author}
  {\bibfnamefont {J.~S.}\ \bibnamefont {Harris}}, \bibinfo {author}
  {\bibfnamefont {K.}~\bibnamefont {Haughian}}, \bibinfo {author}
  {\bibfnamefont {D.}~\bibnamefont {Heinert}}, \bibinfo {author} {\bibfnamefont
  {B.}~\bibnamefont {Lantz}}, \bibinfo {author} {\bibfnamefont {A.~C.}\
  \bibnamefont {Lin}}, \bibinfo {author} {\bibfnamefont {A.~S.}\ \bibnamefont
  {Markosyan}}, \bibinfo {author} {\bibfnamefont {R.}~\bibnamefont {Nawrodt}},
  \bibinfo {author} {\bibfnamefont {R.}~\bibnamefont {Route}},\ and\ \bibinfo
  {author} {\bibfnamefont {S.}~\bibnamefont {Rowan}},\ }\href
  {https://doi.org/10.1088/0264-9381/32/3/035002} {\bibfield  {journal}
  {\bibinfo  {journal} {Classical and Quantum Gravity}\ }\textbf {\bibinfo
  {volume} {32}},\ \bibinfo {pages} {035002} (\bibinfo {year}
  {2015})}\BibitemShut {NoStop}%
\bibitem [{\citenamefont {Winkler}\ \emph {et~al.}(1994)\citenamefont
  {Winkler}, \citenamefont {Rüdiger}, \citenamefont {Schilling}, \citenamefont
  {Strain},\ and\ \citenamefont {Danzmann}}]{Winkler1994}%
  \BibitemOpen
  \bibfield  {author} {\bibinfo {author} {\bibfnamefont {W.}~\bibnamefont
  {Winkler}}, \bibinfo {author} {\bibfnamefont {A.}~\bibnamefont {Rüdiger}},
  \bibinfo {author} {\bibfnamefont {R.}~\bibnamefont {Schilling}}, \bibinfo
  {author} {\bibfnamefont {K.}~\bibnamefont {Strain}},\ and\ \bibinfo {author}
  {\bibfnamefont {K.}~\bibnamefont {Danzmann}},\ }\href
  {https://doi.org/https://doi.org/10.1016/0030-4018(94)90626-2} {\bibfield
  {journal} {\bibinfo  {journal} {Optics Communications}\ }\textbf {\bibinfo
  {volume} {112}},\ \bibinfo {pages} {245} (\bibinfo {year}
  {1994})}\BibitemShut {NoStop}%
\bibitem [{\citenamefont {Somiya}\ \emph {et~al.}(2019)\citenamefont {Somiya},
  \citenamefont {Hirose},\ and\ \citenamefont
  {Michimura}}]{SapphireInhomogeneity}%
  \BibitemOpen
  \bibfield  {author} {\bibinfo {author} {\bibfnamefont {K.}~\bibnamefont
  {Somiya}}, \bibinfo {author} {\bibfnamefont {E.}~\bibnamefont {Hirose}},\
  and\ \bibinfo {author} {\bibfnamefont {Y.}~\bibnamefont {Michimura}},\ }\href
  {https://doi.org/10.1103/PhysRevD.100.082005} {\bibfield  {journal} {\bibinfo
   {journal} {Phys. Rev. D}\ }\textbf {\bibinfo {volume} {100}},\ \bibinfo
  {pages} {082005} (\bibinfo {year} {2019})}\BibitemShut {NoStop}%
\bibitem [{\citenamefont {Hirose}\ \emph {et~al.}(2020)\citenamefont {Hirose},
  \citenamefont {Billingsley}, \citenamefont {Zhang}, \citenamefont {Yamamoto},
  \citenamefont {Pinard}, \citenamefont {Michel}, \citenamefont {Forest},
  \citenamefont {Reichman},\ and\ \citenamefont {Gross}}]{HiroseSapphire}%
  \BibitemOpen
  \bibfield  {author} {\bibinfo {author} {\bibfnamefont {E.}~\bibnamefont
  {Hirose}}, \bibinfo {author} {\bibfnamefont {G.}~\bibnamefont {Billingsley}},
  \bibinfo {author} {\bibfnamefont {L.}~\bibnamefont {Zhang}}, \bibinfo
  {author} {\bibfnamefont {H.}~\bibnamefont {Yamamoto}}, \bibinfo {author}
  {\bibfnamefont {L.}~\bibnamefont {Pinard}}, \bibinfo {author} {\bibfnamefont
  {C.}~\bibnamefont {Michel}}, \bibinfo {author} {\bibfnamefont
  {D.}~\bibnamefont {Forest}}, \bibinfo {author} {\bibfnamefont
  {B.}~\bibnamefont {Reichman}},\ and\ \bibinfo {author} {\bibfnamefont
  {M.}~\bibnamefont {Gross}},\ }\href
  {https://doi.org/10.1103/PhysRevApplied.14.014021} {\bibfield  {journal}
  {\bibinfo  {journal} {Phys. Rev. Appl.}\ }\textbf {\bibinfo {volume} {14}},\
  \bibinfo {pages} {014021} (\bibinfo {year} {2020})}\BibitemShut {NoStop}%
\bibitem [{\citenamefont {Krüger}\ \emph {et~al.}(2015)\citenamefont
  {Krüger}, \citenamefont {Heinert}, \citenamefont {Khalaidovski},
  \citenamefont {Steinlechner}, \citenamefont {Nawrodt}, \citenamefont
  {Schnabel},\ and\ \citenamefont {Lück}}]{AEISiliconBirefringence}%
  \BibitemOpen
  \bibfield  {author} {\bibinfo {author} {\bibfnamefont {C.}~\bibnamefont
  {Krüger}}, \bibinfo {author} {\bibfnamefont {D.}~\bibnamefont {Heinert}},
  \bibinfo {author} {\bibfnamefont {A.}~\bibnamefont {Khalaidovski}}, \bibinfo
  {author} {\bibfnamefont {J.}~\bibnamefont {Steinlechner}}, \bibinfo {author}
  {\bibfnamefont {R.}~\bibnamefont {Nawrodt}}, \bibinfo {author} {\bibfnamefont
  {R.}~\bibnamefont {Schnabel}},\ and\ \bibinfo {author} {\bibfnamefont
  {H.}~\bibnamefont {Lück}},\ }\href
  {https://doi.org/10.1088/0264-9381/33/1/015012} {\bibfield  {journal}
  {\bibinfo  {journal} {Classical and Quantum Gravity}\ }\textbf {\bibinfo
  {volume} {33}},\ \bibinfo {pages} {015012} (\bibinfo {year}
  {2015})}\BibitemShut {NoStop}%
\bibitem [{\citenamefont {Jaberian~Hamedan}\ \emph {et~al.}(2023)\citenamefont
  {Jaberian~Hamedan}, \citenamefont {Adam}, \citenamefont {Blair},
  \citenamefont {Ju},\ and\ \citenamefont {Zhao}}]{UWASiliconBirefringence}%
  \BibitemOpen
  \bibfield  {author} {\bibinfo {author} {\bibfnamefont {V.}~\bibnamefont
  {Jaberian~Hamedan}}, \bibinfo {author} {\bibfnamefont {A.}~\bibnamefont
  {Adam}}, \bibinfo {author} {\bibfnamefont {C.}~\bibnamefont {Blair}},
  \bibinfo {author} {\bibfnamefont {L.}~\bibnamefont {Ju}},\ and\ \bibinfo
  {author} {\bibfnamefont {C.}~\bibnamefont {Zhao}},\ }\href
  {https://doi.org/10.1063/5.0136869} {\bibfield  {journal} {\bibinfo
  {journal} {Applied Physics Letters}\ }\textbf {\bibinfo {volume} {122}},\
  \bibinfo {pages} {064101} (\bibinfo {year} {2023})}\BibitemShut {NoStop}%
\bibitem [{\citenamefont {Winkler}\ \emph {et~al.}(2021)\citenamefont
  {Winkler}, \citenamefont {Perner}, \citenamefont {Truong}, \citenamefont
  {Zhao}, \citenamefont {Bachmann}, \citenamefont {Mayer}, \citenamefont
  {Fellinger}, \citenamefont {Follman}, \citenamefont {Heu}, \citenamefont
  {Deutsch}, \citenamefont {Bailey}, \citenamefont {Peelaers}, \citenamefont
  {Puchegger}, \citenamefont {Fleisher}, \citenamefont {Cole},\ and\
  \citenamefont {Heckl}}]{Winkler2021}%
  \BibitemOpen
  \bibfield  {author} {\bibinfo {author} {\bibfnamefont {G.}~\bibnamefont
  {Winkler}}, \bibinfo {author} {\bibfnamefont {L.~W.}\ \bibnamefont {Perner}},
  \bibinfo {author} {\bibfnamefont {G.-W.}\ \bibnamefont {Truong}}, \bibinfo
  {author} {\bibfnamefont {G.}~\bibnamefont {Zhao}}, \bibinfo {author}
  {\bibfnamefont {D.}~\bibnamefont {Bachmann}}, \bibinfo {author}
  {\bibfnamefont {A.~S.}\ \bibnamefont {Mayer}}, \bibinfo {author}
  {\bibfnamefont {J.}~\bibnamefont {Fellinger}}, \bibinfo {author}
  {\bibfnamefont {D.}~\bibnamefont {Follman}}, \bibinfo {author} {\bibfnamefont
  {P.}~\bibnamefont {Heu}}, \bibinfo {author} {\bibfnamefont {C.}~\bibnamefont
  {Deutsch}}, \bibinfo {author} {\bibfnamefont {D.~M.}\ \bibnamefont {Bailey}},
  \bibinfo {author} {\bibfnamefont {H.}~\bibnamefont {Peelaers}}, \bibinfo
  {author} {\bibfnamefont {S.}~\bibnamefont {Puchegger}}, \bibinfo {author}
  {\bibfnamefont {A.~J.}\ \bibnamefont {Fleisher}}, \bibinfo {author}
  {\bibfnamefont {G.~D.}\ \bibnamefont {Cole}},\ and\ \bibinfo {author}
  {\bibfnamefont {O.~H.}\ \bibnamefont {Heckl}},\ }\href
  {https://doi.org/10.1364/OPTICA.405938} {\bibfield  {journal} {\bibinfo
  {journal} {Optica}\ }\textbf {\bibinfo {volume} {8}},\ \bibinfo {pages} {686}
  (\bibinfo {year} {2021})}\BibitemShut {NoStop}%
\bibitem [{\citenamefont {Tanioka}\ \emph {et~al.}(2023)\citenamefont
  {Tanioka}, \citenamefont {Vander-Hyde}, \citenamefont {Cole}, \citenamefont
  {Penn},\ and\ \citenamefont {Ballmer}}]{Tanioka2023}%
  \BibitemOpen
  \bibfield  {author} {\bibinfo {author} {\bibfnamefont {S.}~\bibnamefont
  {Tanioka}}, \bibinfo {author} {\bibfnamefont {D.}~\bibnamefont
  {Vander-Hyde}}, \bibinfo {author} {\bibfnamefont {G.~D.}\ \bibnamefont
  {Cole}}, \bibinfo {author} {\bibfnamefont {S.~D.}\ \bibnamefont {Penn}},\
  and\ \bibinfo {author} {\bibfnamefont {S.~W.}\ \bibnamefont {Ballmer}},\
  }\href {https://doi.org/10.1103/PhysRevD.107.022003} {\bibfield  {journal}
  {\bibinfo  {journal} {Phys. Rev. D}\ }\textbf {\bibinfo {volume} {107}},\
  \bibinfo {pages} {022003} (\bibinfo {year} {2023})}\BibitemShut {NoStop}%
\bibitem [{\citenamefont {Yu}\ \emph {et~al.}(2023)\citenamefont {Yu},
  \citenamefont {H\"afner}, \citenamefont {Legero}, \citenamefont {Herbers},
  \citenamefont {Nicolodi}, \citenamefont {Ma}, \citenamefont {Riehle},
  \citenamefont {Sterr}, \citenamefont {Kedar}, \citenamefont {Robinson},
  \citenamefont {Oelker},\ and\ \citenamefont {Ye}}]{JILA2022}%
  \BibitemOpen
  \bibfield  {author} {\bibinfo {author} {\bibfnamefont {J.}~\bibnamefont
  {Yu}}, \bibinfo {author} {\bibfnamefont {S.}~\bibnamefont {H\"afner}},
  \bibinfo {author} {\bibfnamefont {T.}~\bibnamefont {Legero}}, \bibinfo
  {author} {\bibfnamefont {S.}~\bibnamefont {Herbers}}, \bibinfo {author}
  {\bibfnamefont {D.}~\bibnamefont {Nicolodi}}, \bibinfo {author}
  {\bibfnamefont {C.~Y.}\ \bibnamefont {Ma}}, \bibinfo {author} {\bibfnamefont
  {F.}~\bibnamefont {Riehle}}, \bibinfo {author} {\bibfnamefont
  {U.}~\bibnamefont {Sterr}}, \bibinfo {author} {\bibfnamefont
  {D.}~\bibnamefont {Kedar}}, \bibinfo {author} {\bibfnamefont {J.~M.}\
  \bibnamefont {Robinson}}, \bibinfo {author} {\bibfnamefont {E.}~\bibnamefont
  {Oelker}},\ and\ \bibinfo {author} {\bibfnamefont {J.}~\bibnamefont {Ye}},\
  }\href {https://doi.org/10.1103/PhysRevX.13.041002} {\bibfield  {journal}
  {\bibinfo  {journal} {Phys. Rev. X}\ }\textbf {\bibinfo {volume} {13}},\
  \bibinfo {pages} {041002} (\bibinfo {year} {2023})}\BibitemShut {NoStop}%
\bibitem [{\citenamefont {Kryhin}\ \emph {et~al.}(2023)\citenamefont {Kryhin},
  \citenamefont {Hall},\ and\ \citenamefont {Sudhir}}]{ThermorefringentNoise}%
  \BibitemOpen
  \bibfield  {author} {\bibinfo {author} {\bibfnamefont {S.}~\bibnamefont
  {Kryhin}}, \bibinfo {author} {\bibfnamefont {E.~D.}\ \bibnamefont {Hall}},\
  and\ \bibinfo {author} {\bibfnamefont {V.}~\bibnamefont {Sudhir}},\ }\href
  {https://doi.org/10.1103/PhysRevD.107.022001} {\bibfield  {journal} {\bibinfo
   {journal} {Phys. Rev. D}\ }\textbf {\bibinfo {volume} {107}},\ \bibinfo
  {pages} {022001} (\bibinfo {year} {2023})}\BibitemShut {NoStop}%
\bibitem [{\citenamefont {Bielsa}\ \emph {et~al.}(2009)\citenamefont {Bielsa},
  \citenamefont {Dupays}, \citenamefont {Fouch{\'e}}, \citenamefont {Battesti},
  \citenamefont {Robilliard},\ and\ \citenamefont {Rizzo}}]{Bielsa2009}%
  \BibitemOpen
  \bibfield  {author} {\bibinfo {author} {\bibfnamefont {F.}~\bibnamefont
  {Bielsa}}, \bibinfo {author} {\bibfnamefont {A.}~\bibnamefont {Dupays}},
  \bibinfo {author} {\bibfnamefont {M.}~\bibnamefont {Fouch{\'e}}}, \bibinfo
  {author} {\bibfnamefont {R.}~\bibnamefont {Battesti}}, \bibinfo {author}
  {\bibfnamefont {C.}~\bibnamefont {Robilliard}},\ and\ \bibinfo {author}
  {\bibfnamefont {C.}~\bibnamefont {Rizzo}},\ }\href
  {https://doi.org/10.1007/s00340-009-3677-7} {\bibfield  {journal} {\bibinfo
  {journal} {Applied Physics B}\ }\textbf {\bibinfo {volume} {97}},\ \bibinfo
  {pages} {457} (\bibinfo {year} {2009})}\BibitemShut {NoStop}%
\bibitem [{\citenamefont {Hollis}\ \emph {et~al.}(2019)\citenamefont {Hollis},
  \citenamefont {Alberts}, \citenamefont {Tanner},\ and\ \citenamefont
  {Mueller}}]{Hollis2019}%
  \BibitemOpen
  \bibfield  {author} {\bibinfo {author} {\bibfnamefont {H.}~\bibnamefont
  {Hollis}}, \bibinfo {author} {\bibfnamefont {G.}~\bibnamefont {Alberts}},
  \bibinfo {author} {\bibfnamefont {D.~B.}\ \bibnamefont {Tanner}},\ and\
  \bibinfo {author} {\bibfnamefont {G.}~\bibnamefont {Mueller}},\ }\href@noop
  {} {\bibinfo {title} {A laser heterodyne polarimeter for birefringence
  measurement}} (\bibinfo {year} {2019}),\ \Eprint
  {https://arxiv.org/abs/1710.03801} {arXiv:1710.03801 [physics.ins-det]}
  \BibitemShut {NoStop}%
\bibitem [{\citenamefont {Zavattini}\ \emph {et~al.}(2018)\citenamefont
  {Zavattini}, \citenamefont {Della~Valle}, \citenamefont {Ejlli},
  \citenamefont {Ni}, \citenamefont {Gastaldi}, \citenamefont {Milotti},
  \citenamefont {Pengo},\ and\ \citenamefont {Ruoso}}]{Zavattini2018}%
  \BibitemOpen
  \bibfield  {author} {\bibinfo {author} {\bibfnamefont {G.}~\bibnamefont
  {Zavattini}}, \bibinfo {author} {\bibfnamefont {F.}~\bibnamefont
  {Della~Valle}}, \bibinfo {author} {\bibfnamefont {A.}~\bibnamefont {Ejlli}},
  \bibinfo {author} {\bibfnamefont {W.-T.}\ \bibnamefont {Ni}}, \bibinfo
  {author} {\bibfnamefont {U.}~\bibnamefont {Gastaldi}}, \bibinfo {author}
  {\bibfnamefont {E.}~\bibnamefont {Milotti}}, \bibinfo {author} {\bibfnamefont
  {R.}~\bibnamefont {Pengo}},\ and\ \bibinfo {author} {\bibfnamefont
  {G.}~\bibnamefont {Ruoso}},\ }\href
  {https://doi.org/10.1140/epjc/s10052-018-6063-y} {\bibfield  {journal}
  {\bibinfo  {journal} {The European Physical Journal C}\ }\textbf {\bibinfo
  {volume} {78}},\ \bibinfo {pages} {585} (\bibinfo {year} {2018})}\BibitemShut
  {NoStop}%
\bibitem [{\citenamefont {Ejlli}\ \emph {et~al.}(2020)\citenamefont {Ejlli},
  \citenamefont {{Della Valle}}, \citenamefont {Gastaldi}, \citenamefont
  {Messineo}, \citenamefont {Pengo}, \citenamefont {Ruoso},\ and\ \citenamefont
  {Zavattini}}]{Ejlli2020}%
  \BibitemOpen
  \bibfield  {author} {\bibinfo {author} {\bibfnamefont {A.}~\bibnamefont
  {Ejlli}}, \bibinfo {author} {\bibfnamefont {F.}~\bibnamefont {{Della
  Valle}}}, \bibinfo {author} {\bibfnamefont {U.}~\bibnamefont {Gastaldi}},
  \bibinfo {author} {\bibfnamefont {G.}~\bibnamefont {Messineo}}, \bibinfo
  {author} {\bibfnamefont {R.}~\bibnamefont {Pengo}}, \bibinfo {author}
  {\bibfnamefont {G.}~\bibnamefont {Ruoso}},\ and\ \bibinfo {author}
  {\bibfnamefont {G.}~\bibnamefont {Zavattini}},\ }\href
  {https://doi.org/https://doi.org/10.1016/j.physrep.2020.06.001} {\bibfield
  {journal} {\bibinfo  {journal} {Physics Reports}\ }\textbf {\bibinfo {volume}
  {871}},\ \bibinfo {pages} {1} (\bibinfo {year} {2020})},\ \bibinfo {note}
  {the PVLAS experiment: A 25 year effort to measure vacuum magnetic
  birefringence}\BibitemShut {NoStop}%
\bibitem [{\citenamefont {Kamioka}(2020)}]{KamiokaDron}%
  \BibitemOpen
  \bibfield  {author} {\bibinfo {author} {\bibfnamefont {S.}~\bibnamefont
  {Kamioka}},\ }\emph {\bibinfo {title} {Search For Vacuum Magnetic
  Birefringence with a high repetitive pulsed magnet}},\ \href
  {https://tabletop.icepp.s.u-tokyo.ac.jp/wp-content/uploads/2021/02/Dron-kamioka.pdf}
  {Ph.D. thesis},\ \bibinfo  {school} {University of Tokyo} (\bibinfo {year}
  {2020})\BibitemShut {NoStop}%
\bibitem [{\citenamefont {Obata}\ \emph {et~al.}(2018)\citenamefont {Obata},
  \citenamefont {Fujita},\ and\ \citenamefont {Michimura}}]{DANCE}%
  \BibitemOpen
  \bibfield  {author} {\bibinfo {author} {\bibfnamefont {I.}~\bibnamefont
  {Obata}}, \bibinfo {author} {\bibfnamefont {T.}~\bibnamefont {Fujita}},\ and\
  \bibinfo {author} {\bibfnamefont {Y.}~\bibnamefont {Michimura}},\ }\href
  {https://doi.org/10.1103/PhysRevLett.121.161301} {\bibfield  {journal}
  {\bibinfo  {journal} {Phys. Rev. Lett.}\ }\textbf {\bibinfo {volume} {121}},\
  \bibinfo {pages} {161301} (\bibinfo {year} {2018})}\BibitemShut {NoStop}%
\bibitem [{\citenamefont {Liu}\ \emph {et~al.}(2019)\citenamefont {Liu},
  \citenamefont {Elwood}, \citenamefont {Evans},\ and\ \citenamefont
  {Thaler}}]{ADBC}%
  \BibitemOpen
  \bibfield  {author} {\bibinfo {author} {\bibfnamefont {H.}~\bibnamefont
  {Liu}}, \bibinfo {author} {\bibfnamefont {B.~D.}\ \bibnamefont {Elwood}},
  \bibinfo {author} {\bibfnamefont {M.}~\bibnamefont {Evans}},\ and\ \bibinfo
  {author} {\bibfnamefont {J.}~\bibnamefont {Thaler}},\ }\href
  {https://doi.org/10.1103/PhysRevD.100.023548} {\bibfield  {journal} {\bibinfo
   {journal} {Phys. Rev. D}\ }\textbf {\bibinfo {volume} {100}},\ \bibinfo
  {pages} {023548} (\bibinfo {year} {2019})}\BibitemShut {NoStop}%
\bibitem [{\citenamefont {Martynov}\ and\ \citenamefont {Miao}(2020)}]{LIDA}%
  \BibitemOpen
  \bibfield  {author} {\bibinfo {author} {\bibfnamefont {D.}~\bibnamefont
  {Martynov}}\ and\ \bibinfo {author} {\bibfnamefont {H.}~\bibnamefont
  {Miao}},\ }\href {https://doi.org/10.1103/PhysRevD.101.095034} {\bibfield
  {journal} {\bibinfo  {journal} {Phys. Rev. D}\ }\textbf {\bibinfo {volume}
  {101}},\ \bibinfo {pages} {095034} (\bibinfo {year} {2020})}\BibitemShut
  {NoStop}%
\bibitem [{\citenamefont {Nagano}\ \emph {et~al.}(2019)\citenamefont {Nagano},
  \citenamefont {Fujita}, \citenamefont {Michimura},\ and\ \citenamefont
  {Obata}}]{ADAM-GD}%
  \BibitemOpen
  \bibfield  {author} {\bibinfo {author} {\bibfnamefont {K.}~\bibnamefont
  {Nagano}}, \bibinfo {author} {\bibfnamefont {T.}~\bibnamefont {Fujita}},
  \bibinfo {author} {\bibfnamefont {Y.}~\bibnamefont {Michimura}},\ and\
  \bibinfo {author} {\bibfnamefont {I.}~\bibnamefont {Obata}},\ }\href
  {https://doi.org/10.1103/PhysRevLett.123.111301} {\bibfield  {journal}
  {\bibinfo  {journal} {Phys. Rev. Lett.}\ }\textbf {\bibinfo {volume} {123}},\
  \bibinfo {pages} {111301} (\bibinfo {year} {2019})}\BibitemShut {NoStop}%
\bibitem [{\citenamefont {Nagano}\ \emph {et~al.}(2021)\citenamefont {Nagano},
  \citenamefont {Nakatsuka}, \citenamefont {Morisaki}, \citenamefont {Fujita},
  \citenamefont {Michimura},\ and\ \citenamefont {Obata}}]{ADAM-GDTR}%
  \BibitemOpen
  \bibfield  {author} {\bibinfo {author} {\bibfnamefont {K.}~\bibnamefont
  {Nagano}}, \bibinfo {author} {\bibfnamefont {H.}~\bibnamefont {Nakatsuka}},
  \bibinfo {author} {\bibfnamefont {S.}~\bibnamefont {Morisaki}}, \bibinfo
  {author} {\bibfnamefont {T.}~\bibnamefont {Fujita}}, \bibinfo {author}
  {\bibfnamefont {Y.}~\bibnamefont {Michimura}},\ and\ \bibinfo {author}
  {\bibfnamefont {I.}~\bibnamefont {Obata}},\ }\href
  {https://doi.org/10.1103/PhysRevD.104.062008} {\bibfield  {journal} {\bibinfo
   {journal} {Phys. Rev. D}\ }\textbf {\bibinfo {volume} {104}},\ \bibinfo
  {pages} {062008} (\bibinfo {year} {2021})}\BibitemShut {NoStop}%
\bibitem [{\citenamefont {Oelker}\ \emph {et~al.}(2016)\citenamefont {Oelker},
  \citenamefont {Mansell}, \citenamefont {Tse}, \citenamefont {Miller},
  \citenamefont {Matichard}, \citenamefont {Barsotti}, \citenamefont
  {Fritschel}, \citenamefont {McClelland}, \citenamefont {Evans},\ and\
  \citenamefont {Mavalvala}}]{Oelker2016}%
  \BibitemOpen
  \bibfield  {author} {\bibinfo {author} {\bibfnamefont {E.}~\bibnamefont
  {Oelker}}, \bibinfo {author} {\bibfnamefont {G.}~\bibnamefont {Mansell}},
  \bibinfo {author} {\bibfnamefont {M.}~\bibnamefont {Tse}}, \bibinfo {author}
  {\bibfnamefont {J.}~\bibnamefont {Miller}}, \bibinfo {author} {\bibfnamefont
  {F.}~\bibnamefont {Matichard}}, \bibinfo {author} {\bibfnamefont
  {L.}~\bibnamefont {Barsotti}}, \bibinfo {author} {\bibfnamefont
  {P.}~\bibnamefont {Fritschel}}, \bibinfo {author} {\bibfnamefont {D.~E.}\
  \bibnamefont {McClelland}}, \bibinfo {author} {\bibfnamefont
  {M.}~\bibnamefont {Evans}},\ and\ \bibinfo {author} {\bibfnamefont
  {N.}~\bibnamefont {Mavalvala}},\ }\href
  {https://doi.org/10.1364/OPTICA.3.000682} {\bibfield  {journal} {\bibinfo
  {journal} {Optica}\ }\textbf {\bibinfo {volume} {3}},\ \bibinfo {pages} {682}
  (\bibinfo {year} {2016})}\BibitemShut {NoStop}%
\bibitem [{\citenamefont {Jones}(1941)}]{JonesCalculus}%
  \BibitemOpen
  \bibfield  {author} {\bibinfo {author} {\bibfnamefont {R.~C.}\ \bibnamefont
  {Jones}},\ }\href {https://doi.org/10.1364/JOSA.31.000488} {\bibfield
  {journal} {\bibinfo  {journal} {J. Opt. Soc. Am.}\ }\textbf {\bibinfo
  {volume} {31}},\ \bibinfo {pages} {488} (\bibinfo {year} {1941})}\BibitemShut
  {NoStop}%
\bibitem [{\citenamefont {Brandi}\ \emph {et~al.}(1997)\citenamefont {Brandi},
  \citenamefont {Della~Valle}, \citenamefont {De~Riva}, \citenamefont
  {Micossi}, \citenamefont {Perrone}, \citenamefont {Rizzo}, \citenamefont
  {Ruoso},\ and\ \citenamefont {Zavattini}}]{Brandi1997}%
  \BibitemOpen
  \bibfield  {author} {\bibinfo {author} {\bibfnamefont {F.}~\bibnamefont
  {Brandi}}, \bibinfo {author} {\bibfnamefont {F.}~\bibnamefont {Della~Valle}},
  \bibinfo {author} {\bibfnamefont {A.~M.}\ \bibnamefont {De~Riva}}, \bibinfo
  {author} {\bibfnamefont {P.}~\bibnamefont {Micossi}}, \bibinfo {author}
  {\bibfnamefont {F.}~\bibnamefont {Perrone}}, \bibinfo {author} {\bibfnamefont
  {C.}~\bibnamefont {Rizzo}}, \bibinfo {author} {\bibfnamefont
  {G.}~\bibnamefont {Ruoso}},\ and\ \bibinfo {author} {\bibfnamefont
  {G.}~\bibnamefont {Zavattini}},\ }\href
  {https://doi.org/10.1007/s003400050283} {\bibfield  {journal} {\bibinfo
  {journal} {Applied Physics B}\ }\textbf {\bibinfo {volume} {65}},\ \bibinfo
  {pages} {351} (\bibinfo {year} {1997})}\BibitemShut {NoStop}%
\bibitem [{\citenamefont {Miller}\ \emph {et~al.}(2015)\citenamefont {Miller},
  \citenamefont {Barsotti}, \citenamefont {Vitale}, \citenamefont {Fritschel},
  \citenamefont {Evans},\ and\ \citenamefont {Sigg}}]{Aplus}%
  \BibitemOpen
  \bibfield  {author} {\bibinfo {author} {\bibfnamefont {J.}~\bibnamefont
  {Miller}}, \bibinfo {author} {\bibfnamefont {L.}~\bibnamefont {Barsotti}},
  \bibinfo {author} {\bibfnamefont {S.}~\bibnamefont {Vitale}}, \bibinfo
  {author} {\bibfnamefont {P.}~\bibnamefont {Fritschel}}, \bibinfo {author}
  {\bibfnamefont {M.}~\bibnamefont {Evans}},\ and\ \bibinfo {author}
  {\bibfnamefont {D.}~\bibnamefont {Sigg}},\ }\href
  {https://doi.org/10.1103/PhysRevD.91.062005} {\bibfield  {journal} {\bibinfo
  {journal} {Phys. Rev. D}\ }\textbf {\bibinfo {volume} {91}},\ \bibinfo
  {pages} {062005} (\bibinfo {year} {2015})}\BibitemShut {NoStop}%
\bibitem [{\citenamefont {Evans}\ \emph {et~al.}(2020)\citenamefont {Evans},
  \citenamefont {Sturani}, \citenamefont {Vitale},\ and\ \citenamefont
  {Hall}}]{LIGO-T1500293}%
  \BibitemOpen
  \bibfield  {author} {\bibinfo {author} {\bibfnamefont {M.}~\bibnamefont
  {Evans}}, \bibinfo {author} {\bibfnamefont {R.}~\bibnamefont {Sturani}},
  \bibinfo {author} {\bibfnamefont {S.}~\bibnamefont {Vitale}},\ and\ \bibinfo
  {author} {\bibfnamefont {E.}~\bibnamefont {Hall}},\ }\href
  {https://dcc.ligo.org/LIGO-T1500293-v13/public} {\emph {\bibinfo {title}
  {Unofficial sensitivity curves (ASD) for aLIGO, Kagra, Virgo, Voyager, Cosmic
  Explorer, and Einstein Telescope}}},\ \bibinfo {type} {Tech. Rep.}\ \bibinfo
  {number} {LIGO-T1500293-v13}\ (\bibinfo {year} {2020})\BibitemShut {NoStop}%
\bibitem [{\citenamefont {Kuns}\ \emph {et~al.}(2023)\citenamefont {Kuns},
  \citenamefont {Hall}, \citenamefont {Srivastava}, \citenamefont {Send},
  \citenamefont {Evans}, \citenamefont {Fritschel}, \citenamefont {McCuller},
  \citenamefont {Wipf},\ and\ \citenamefont {Ballmer}}]{CE-T2000017}%
  \BibitemOpen
  \bibfield  {author} {\bibinfo {author} {\bibfnamefont {K.}~\bibnamefont
  {Kuns}}, \bibinfo {author} {\bibfnamefont {E.}~\bibnamefont {Hall}}, \bibinfo
  {author} {\bibfnamefont {V.}~\bibnamefont {Srivastava}}, \bibinfo {author}
  {\bibfnamefont {J.~S.}\ \bibnamefont {Send}}, \bibinfo {author}
  {\bibfnamefont {M.}~\bibnamefont {Evans}}, \bibinfo {author} {\bibfnamefont
  {P.}~\bibnamefont {Fritschel}}, \bibinfo {author} {\bibfnamefont
  {L.}~\bibnamefont {McCuller}}, \bibinfo {author} {\bibfnamefont
  {C.}~\bibnamefont {Wipf}},\ and\ \bibinfo {author} {\bibfnamefont
  {S.}~\bibnamefont {Ballmer}},\ }\href
  {https://dcc.cosmicexplorer.org/CE-T2000017/public} {\emph {\bibinfo {title}
  {Cosmic Explorer Strain Sensitivity}}},\ \bibinfo {type} {Tech. Rep.}\
  \bibinfo {number} {CE-T2000017-v6}\ (\bibinfo {year} {2023})\BibitemShut
  {NoStop}%
\bibitem [{\citenamefont {{ET~design~team}}(2018)}]{ET-0000A-18}%
  \BibitemOpen
  \bibfield  {author} {\bibinfo {author} {\bibnamefont {{ET~design~team}}},\
  }\href {https://tds.virgo-gw.eu/ql/?c=12989} {\emph {\bibinfo {title} {ET-D
  sensitivity curve txt file}}},\ \bibinfo {type} {Tech. Rep.}\ \bibinfo
  {number} {ET-0000A-18}\ (\bibinfo {year} {2018})\BibitemShut {NoStop}%
\bibitem [{\citenamefont {Essick}\ \emph {et~al.}(2017)\citenamefont {Essick},
  \citenamefont {Vitale},\ and\ \citenamefont {Evans}}]{Essick2017}%
  \BibitemOpen
  \bibfield  {author} {\bibinfo {author} {\bibfnamefont {R.}~\bibnamefont
  {Essick}}, \bibinfo {author} {\bibfnamefont {S.}~\bibnamefont {Vitale}},\
  and\ \bibinfo {author} {\bibfnamefont {M.}~\bibnamefont {Evans}},\ }\href
  {https://doi.org/10.1103/PhysRevD.96.084004} {\bibfield  {journal} {\bibinfo
  {journal} {Phys. Rev. D}\ }\textbf {\bibinfo {volume} {96}},\ \bibinfo
  {pages} {084004} (\bibinfo {year} {2017})}\BibitemShut {NoStop}%
\bibitem [{\citenamefont {Aston}\ \emph {et~al.}(2012)\citenamefont {Aston},
  \citenamefont {Barton}, \citenamefont {Bell}, \citenamefont {Beveridge},
  \citenamefont {Bland}, \citenamefont {Brummitt}, \citenamefont {Cagnoli},
  \citenamefont {Cantley}, \citenamefont {Carbone}, \citenamefont {Cumming},
  \citenamefont {Cunningham}, \citenamefont {Cutler}, \citenamefont
  {Greenhalgh}, \citenamefont {Hammond}, \citenamefont {Haughian},
  \citenamefont {Hayler}, \citenamefont {Heptonstall}, \citenamefont {Heefner},
  \citenamefont {Hoyland}, \citenamefont {Hough}, \citenamefont {Jones},
  \citenamefont {Kissel}, \citenamefont {Kumar}, \citenamefont {Lockerbie},
  \citenamefont {Lodhia}, \citenamefont {Martin}, \citenamefont {Murray},
  \citenamefont {O’Dell}, \citenamefont {Plissi}, \citenamefont {Reid},
  \citenamefont {Romie}, \citenamefont {Robertson}, \citenamefont {Rowan},
  \citenamefont {Shapiro}, \citenamefont {Speake}, \citenamefont {Strain},
  \citenamefont {Tokmakov}, \citenamefont {Torrie}, \citenamefont {van Veggel},
  \citenamefont {Vecchio},\ and\ \citenamefont {Wilmut}}]{aLIGOQuad}%
  \BibitemOpen
  \bibfield  {author} {\bibinfo {author} {\bibfnamefont {S.~M.}\ \bibnamefont
  {Aston}}, \bibinfo {author} {\bibfnamefont {M.~A.}\ \bibnamefont {Barton}},
  \bibinfo {author} {\bibfnamefont {A.~S.}\ \bibnamefont {Bell}}, \bibinfo
  {author} {\bibfnamefont {N.}~\bibnamefont {Beveridge}}, \bibinfo {author}
  {\bibfnamefont {B.}~\bibnamefont {Bland}}, \bibinfo {author} {\bibfnamefont
  {A.~J.}\ \bibnamefont {Brummitt}}, \bibinfo {author} {\bibfnamefont
  {G.}~\bibnamefont {Cagnoli}}, \bibinfo {author} {\bibfnamefont {C.~A.}\
  \bibnamefont {Cantley}}, \bibinfo {author} {\bibfnamefont {L.}~\bibnamefont
  {Carbone}}, \bibinfo {author} {\bibfnamefont {A.~V.}\ \bibnamefont
  {Cumming}}, \bibinfo {author} {\bibfnamefont {L.}~\bibnamefont {Cunningham}},
  \bibinfo {author} {\bibfnamefont {R.~M.}\ \bibnamefont {Cutler}}, \bibinfo
  {author} {\bibfnamefont {R.~J.~S.}\ \bibnamefont {Greenhalgh}}, \bibinfo
  {author} {\bibfnamefont {G.~D.}\ \bibnamefont {Hammond}}, \bibinfo {author}
  {\bibfnamefont {K.}~\bibnamefont {Haughian}}, \bibinfo {author}
  {\bibfnamefont {T.~M.}\ \bibnamefont {Hayler}}, \bibinfo {author}
  {\bibfnamefont {A.}~\bibnamefont {Heptonstall}}, \bibinfo {author}
  {\bibfnamefont {J.}~\bibnamefont {Heefner}}, \bibinfo {author} {\bibfnamefont
  {D.}~\bibnamefont {Hoyland}}, \bibinfo {author} {\bibfnamefont
  {J.}~\bibnamefont {Hough}}, \bibinfo {author} {\bibfnamefont
  {R.}~\bibnamefont {Jones}}, \bibinfo {author} {\bibfnamefont {J.~S.}\
  \bibnamefont {Kissel}}, \bibinfo {author} {\bibfnamefont {R.}~\bibnamefont
  {Kumar}}, \bibinfo {author} {\bibfnamefont {N.~A.}\ \bibnamefont
  {Lockerbie}}, \bibinfo {author} {\bibfnamefont {D.}~\bibnamefont {Lodhia}},
  \bibinfo {author} {\bibfnamefont {I.~W.}\ \bibnamefont {Martin}}, \bibinfo
  {author} {\bibfnamefont {P.~G.}\ \bibnamefont {Murray}}, \bibinfo {author}
  {\bibfnamefont {J.}~\bibnamefont {O’Dell}}, \bibinfo {author}
  {\bibfnamefont {M.~V.}\ \bibnamefont {Plissi}}, \bibinfo {author}
  {\bibfnamefont {S.}~\bibnamefont {Reid}}, \bibinfo {author} {\bibfnamefont
  {J.}~\bibnamefont {Romie}}, \bibinfo {author} {\bibfnamefont {N.~A.}\
  \bibnamefont {Robertson}}, \bibinfo {author} {\bibfnamefont {S.}~\bibnamefont
  {Rowan}}, \bibinfo {author} {\bibfnamefont {B.}~\bibnamefont {Shapiro}},
  \bibinfo {author} {\bibfnamefont {C.~C.}\ \bibnamefont {Speake}}, \bibinfo
  {author} {\bibfnamefont {K.~A.}\ \bibnamefont {Strain}}, \bibinfo {author}
  {\bibfnamefont {K.~V.}\ \bibnamefont {Tokmakov}}, \bibinfo {author}
  {\bibfnamefont {C.}~\bibnamefont {Torrie}}, \bibinfo {author} {\bibfnamefont
  {A.~A.}\ \bibnamefont {van Veggel}}, \bibinfo {author} {\bibfnamefont
  {A.}~\bibnamefont {Vecchio}},\ and\ \bibinfo {author} {\bibfnamefont
  {I.}~\bibnamefont {Wilmut}},\ }\href
  {https://doi.org/10.1088/0264-9381/29/23/235004} {\bibfield  {journal}
  {\bibinfo  {journal} {Classical and Quantum Gravity}\ }\textbf {\bibinfo
  {volume} {29}},\ \bibinfo {pages} {235004} (\bibinfo {year}
  {2012})}\BibitemShut {NoStop}%
\bibitem [{\citenamefont {Coughlin}\ \emph {et~al.}(2018)\citenamefont
  {Coughlin}, \citenamefont {Cirone}, \citenamefont {Meyers}, \citenamefont
  {Atsuta}, \citenamefont {Boschi}, \citenamefont {Chincarini}, \citenamefont
  {Christensen}, \citenamefont {De~Rosa}, \citenamefont {Effler}, \citenamefont
  {Fiori}, \citenamefont {Go\l{}kowski}, \citenamefont {Guidry}, \citenamefont
  {Harms}, \citenamefont {Hayama}, \citenamefont {Kataoka}, \citenamefont
  {Kubisz}, \citenamefont {Kulak}, \citenamefont {Laxen}, \citenamefont
  {Matas}, \citenamefont {Mlynarczyk}, \citenamefont {Ogawa}, \citenamefont
  {Paoletti}, \citenamefont {Salvador}, \citenamefont {Schofield},
  \citenamefont {Somiya},\ and\ \citenamefont {Thrane}}]{Coughlin2018}%
  \BibitemOpen
  \bibfield  {author} {\bibinfo {author} {\bibfnamefont {M.~W.}\ \bibnamefont
  {Coughlin}}, \bibinfo {author} {\bibfnamefont {A.}~\bibnamefont {Cirone}},
  \bibinfo {author} {\bibfnamefont {P.}~\bibnamefont {Meyers}}, \bibinfo
  {author} {\bibfnamefont {S.}~\bibnamefont {Atsuta}}, \bibinfo {author}
  {\bibfnamefont {V.}~\bibnamefont {Boschi}}, \bibinfo {author} {\bibfnamefont
  {A.}~\bibnamefont {Chincarini}}, \bibinfo {author} {\bibfnamefont {N.~L.}\
  \bibnamefont {Christensen}}, \bibinfo {author} {\bibfnamefont
  {R.}~\bibnamefont {De~Rosa}}, \bibinfo {author} {\bibfnamefont
  {A.}~\bibnamefont {Effler}}, \bibinfo {author} {\bibfnamefont
  {I.}~\bibnamefont {Fiori}}, \bibinfo {author} {\bibfnamefont
  {M.}~\bibnamefont {Go\l{}kowski}}, \bibinfo {author} {\bibfnamefont
  {M.}~\bibnamefont {Guidry}}, \bibinfo {author} {\bibfnamefont
  {J.}~\bibnamefont {Harms}}, \bibinfo {author} {\bibfnamefont
  {K.}~\bibnamefont {Hayama}}, \bibinfo {author} {\bibfnamefont
  {Y.}~\bibnamefont {Kataoka}}, \bibinfo {author} {\bibfnamefont
  {J.}~\bibnamefont {Kubisz}}, \bibinfo {author} {\bibfnamefont
  {A.}~\bibnamefont {Kulak}}, \bibinfo {author} {\bibfnamefont
  {M.}~\bibnamefont {Laxen}}, \bibinfo {author} {\bibfnamefont
  {A.}~\bibnamefont {Matas}}, \bibinfo {author} {\bibfnamefont
  {J.}~\bibnamefont {Mlynarczyk}}, \bibinfo {author} {\bibfnamefont
  {T.}~\bibnamefont {Ogawa}}, \bibinfo {author} {\bibfnamefont
  {F.}~\bibnamefont {Paoletti}}, \bibinfo {author} {\bibfnamefont
  {J.}~\bibnamefont {Salvador}}, \bibinfo {author} {\bibfnamefont
  {R.}~\bibnamefont {Schofield}}, \bibinfo {author} {\bibfnamefont
  {K.}~\bibnamefont {Somiya}},\ and\ \bibinfo {author} {\bibfnamefont
  {E.}~\bibnamefont {Thrane}},\ }\href
  {https://doi.org/10.1103/PhysRevD.97.102007} {\bibfield  {journal} {\bibinfo
  {journal} {Phys. Rev. D}\ }\textbf {\bibinfo {volume} {97}},\ \bibinfo
  {pages} {102007} (\bibinfo {year} {2018})}\BibitemShut {NoStop}%
\bibitem [{\citenamefont {Snetkov}\ and\ \citenamefont
  {Yakovlev}(2022)}]{SiVerdet}%
  \BibitemOpen
  \bibfield  {author} {\bibinfo {author} {\bibfnamefont {I.}~\bibnamefont
  {Snetkov}}\ and\ \bibinfo {author} {\bibfnamefont {A.}~\bibnamefont
  {Yakovlev}},\ }\href {https://doi.org/10.1364/OL.452218} {\bibfield
  {journal} {\bibinfo  {journal} {Opt. Lett.}\ }\textbf {\bibinfo {volume}
  {47}},\ \bibinfo {pages} {1895} (\bibinfo {year} {2022})}\BibitemShut
  {NoStop}%
\bibitem [{\citenamefont {Fricke}\ \emph {et~al.}(2012)\citenamefont {Fricke},
  \citenamefont {Smith-Lefebvre}, \citenamefont {Abbott}, \citenamefont
  {Adhikari}, \citenamefont {Dooley}, \citenamefont {Evans}, \citenamefont
  {Fritschel}, \citenamefont {Frolov}, \citenamefont {Kawabe}, \citenamefont
  {Kissel}, \citenamefont {Slagmolen},\ and\ \citenamefont
  {Waldman}}]{DCreadout}%
  \BibitemOpen
  \bibfield  {author} {\bibinfo {author} {\bibfnamefont {T.~T.}\ \bibnamefont
  {Fricke}}, \bibinfo {author} {\bibfnamefont {N.~D.}\ \bibnamefont
  {Smith-Lefebvre}}, \bibinfo {author} {\bibfnamefont {R.}~\bibnamefont
  {Abbott}}, \bibinfo {author} {\bibfnamefont {R.}~\bibnamefont {Adhikari}},
  \bibinfo {author} {\bibfnamefont {K.~L.}\ \bibnamefont {Dooley}}, \bibinfo
  {author} {\bibfnamefont {M.}~\bibnamefont {Evans}}, \bibinfo {author}
  {\bibfnamefont {P.}~\bibnamefont {Fritschel}}, \bibinfo {author}
  {\bibfnamefont {V.~V.}\ \bibnamefont {Frolov}}, \bibinfo {author}
  {\bibfnamefont {K.}~\bibnamefont {Kawabe}}, \bibinfo {author} {\bibfnamefont
  {J.~S.}\ \bibnamefont {Kissel}}, \bibinfo {author} {\bibfnamefont {B.~J.~J.}\
  \bibnamefont {Slagmolen}},\ and\ \bibinfo {author} {\bibfnamefont {S.~J.}\
  \bibnamefont {Waldman}},\ }\href
  {https://doi.org/10.1088/0264-9381/29/6/065005} {\bibfield  {journal}
  {\bibinfo  {journal} {Classical and Quantum Gravity}\ }\textbf {\bibinfo
  {volume} {29}},\ \bibinfo {pages} {065005} (\bibinfo {year}
  {2012})}\BibitemShut {NoStop}%
\bibitem [{\citenamefont {Lawrence}(2003)}]{Lawrence}%
  \BibitemOpen
  \bibfield  {author} {\bibinfo {author} {\bibfnamefont {R.~C.}\ \bibnamefont
  {Lawrence}},\ }\emph {\bibinfo {title} {Active wavefront correction in laser
  interferometric gravitational wave detectors}},\ \href
  {https://dspace.mit.edu/handle/1721.1/29308} {Ph.D. thesis},\ \bibinfo
  {school} {Massachusetts Institute of Technology} (\bibinfo {year}
  {2003})\BibitemShut {NoStop}%
\bibitem [{\citenamefont {Tokunari}\ \emph {et~al.}(2006)\citenamefont
  {Tokunari}, \citenamefont {Hayakawa}, \citenamefont {Yamamoto}, \citenamefont
  {Uchiyama}, \citenamefont {Miyoki}, \citenamefont {Ohashi},\ and\
  \citenamefont {Kuroda}}]{Tokunari2006}%
  \BibitemOpen
  \bibfield  {author} {\bibinfo {author} {\bibfnamefont {M.}~\bibnamefont
  {Tokunari}}, \bibinfo {author} {\bibfnamefont {H.}~\bibnamefont {Hayakawa}},
  \bibinfo {author} {\bibfnamefont {K.}~\bibnamefont {Yamamoto}}, \bibinfo
  {author} {\bibfnamefont {T.}~\bibnamefont {Uchiyama}}, \bibinfo {author}
  {\bibfnamefont {S.}~\bibnamefont {Miyoki}}, \bibinfo {author} {\bibfnamefont
  {M.}~\bibnamefont {Ohashi}},\ and\ \bibinfo {author} {\bibfnamefont
  {K.}~\bibnamefont {Kuroda}},\ }\href
  {https://doi.org/10.1088/1742-6596/32/1/066} {\bibfield  {journal} {\bibinfo
  {journal} {Journal of Physics: Conference Series}\ }\textbf {\bibinfo
  {volume} {32}},\ \bibinfo {pages} {432} (\bibinfo {year} {2006})}\BibitemShut
  {NoStop}%
\end{thebibliography}%
\end{document}